\documentclass[12pt]{article}
\usepackage{graphicx}
\usepackage{url}
\usepackage{cite}
\usepackage[margin=1.25 in]{geometry}
\usepackage[colorlinks = true, linkcolor = blue, urlcolor = blue,
      citecolor = blue, anchorcolor = blue]{hyperref}

\newcommand\pubnumber{SLAC-PUB-17517}
\newcommand\pubdate{March 2020}

\def\SLAC{SLAC,
    Stanford University, Menlo Park, California 94025 USA}
\def\doeack{\footnote{Work supported by the US Department of Energy,
                     contract DE--AC02--76SF00515.}}

\def\Title#1{\begin{center} {\Large #1 } \end{center}}
\def\Author#1{\begin{center}{ \sc #1} \end{center}}

\newcommand\pubblock{\rightline{\begin{tabular}{l} \pubnumber\\
         \pubdate \end{tabular}}}
\newenvironment{Abstract}{\begin{quotation} \begin{center}
                       ABSTRACT
     \end{center}\bigskip  }{\end{quotation}}
\newenvironment{Presented}{\begin{quotation} \begin{center} 
             CONTRIBUTED TO\end{center}\bigskip 
      \begin{center}\begin{large}}{\end{large}\end{center} \end{quotation}}

\def\Acknowledgements{\bigskip  \bigskip \begin{center} \begin{large}
             \bf ACKNOWLEDGEMENTS \end{large}\end{center}}
\def\beq{\begin{equation}}
\def\eeq#1{\label{#1}\end{equation}}
\def\eeqn{\end{equation}}

\newenvironment{Eqnarray}%
   {\arraycolsep 0.14em\begin{eqnarray}}{\end{eqnarray}}
\def\beqa{\begin{Eqnarray}}
\def\eeqa#1{\label{#1}\end{Eqnarray}}
\def\eeqan{\end{Eqnarray}}
\def\CR{\nonumber \\ }

\def\leqn#1{(\ref{#1})}

\let\bar=\overbar

\def\etal{{\it et al.}}

\def\lsim{\mathrel{\raise.3ex\hbox{$<$\kern-.75em\lower1ex\hbox{$\sim$}}}}
\def\gsim{\mathrel{\raise.3ex\hbox{$>$\kern-.75em\lower1ex\hbox{$\sim$}}}}

\def\D{{\cal D}}
\def\L{{\cal L}}

\def\half{\frac{1}{2}}

\def\del{\partial}
\def\Dslash{\not{\hbox{\kern-4pt $D$}}}
\def\dslash{\not{\hbox{\kern-2pt $\del$}}}

\def\Dlr{\mathrel{\raise1.5ex\hbox{$\leftrightarrow$\kern-1em\lower1.5ex\hbox{$D$}}}}

\def\ee{e^+e^-}
\def\sstw{\sin^2\theta_w}

\def\msb{{\bar{\scriptsize M \kern -1pt S}}}

\def\drb{{\bar{\scriptsize D \kern -1pt R}}}

\makeatletter
\def\section{\@startsection{section}{0}{\z@}{5.5ex plus .5ex minus
 1.5ex}{2.3ex plus .2ex}{\large\bf}}
\def\subsection{\@startsection{subsection}{1}{\z@}{3.5ex plus .5ex minus
 1.5ex}{1.3ex plus .2ex}{\normalsize\bf}}
\def\subsubsection{\@startsection{subsubsection}{2}{\z@}{-3.5ex plus
-1ex minus  -.2ex}{2.3ex plus .2ex}{\normalsize\sl}}

\renewcommand{\@makecaption}[2]{%
   \vskip 10pt
   \setbox\@tempboxa\hbox{\small #1: #2}
   \ifdim \wd\@tempboxa >\hsize     % IF longer than one line:
       \small #1: #2\par          %   THEN set as ordinary paragraph.
     \else                        %   ELSE  center.
       \hbox to\hsize{\hfil\box\@tempboxa\hfil}
   \fi}

\makeatother

\begin{document}
\begin{titlepage}
\pubblock

\vfill
\Title{Searching for New Physics using\CR  Precision Standard Model Measurements}
\vfill
\Author{ Michael E. Peskin\doeack}
 \medskip
\begin{center} 
\SLAC  
\end{center}
\vfill
\begin{Abstract}
New physics interactions beyond the Standard Model can make themselves
known as small corrections to Standard Model reactions.  There is a
diverse array of proposals for new physics, and so any parametrization
of those effects must be as general and all-inclusive as possible.
This can be accomplished by the use of Standard Model Effective Field
Theory  (SMEFT).   In this article, part of the celebration of 50
years of the Standard Model of particle physics, I describe how SMEFT
has been applied to search for new physics in fermion-fermion
scattering and precision electroweak analysis and how it will be
applied in the precision study of the Higgs boson.
\end{Abstract} 
\vfill
\begin{Presented}
  The Standard Model at 50 Years\\
  Case Western Reserve University,  June 1-4, 2018
  \end{Presented}
\vfill
\end{titlepage}

\hbox to \hsize{\null}

%\newpage

\tableofcontents

\def\thefootnote{\fnsymbol{footnote}}
\newpage
\setcounter{page}{1}

\setcounter{footnote}{0}

\section{Introduction}

As has been well documented at this symposium, the Standard Model has
been remarkably successful at explaining a wide range of experimental
measurements.   From low-energy observables in
weak interaction decays to multiparticle production at the highest
energies of the LHC, the Standard Model seems to give a complete
description of the reactions of elementary particles.

Still, there are good reasons to believe that the Standard Model is an
incomplete description of nature, and that additional fundamental
interactions are waiting to be discovered.

There are many methods to search for these new interactions.   One way
is to search for new elementary particles that can be produced at high
energies.
Another way is to search in low-energy processes for specific
interactions, for example, flavor- or CP-violating, that are forbidden
in the Standard Model.  A third way is to use our ability to perform
high-precision calculations  in the electroweak sectors of the
Standard Model to search for small deviations from those predictions.
This last method can be sensitive to new interactions well above the
accelerator center of mass energy.   It can also be remarkably robust,
sensitive to a very wide variety of models.

To understand the constraints that come from precision Standard Model
tests, it is useful to have a formalism that can describe as large a
range of new physics models as possible.   This is supplied by
Standard Model Effective Field Theory (SMEFT).   In this article, I
will review some of the applications of SMEFT to the interpretation of
precision measurements.

Recently, Brivio and Trott have given a comprehensive review of
SMEFT~\cite{Brivio}.   In this article, I will have relatively little
to say about the formalism of SMEFT, its renormalization, and the
computation of loop corrections in this framework.  Instead, I will
emphasize its practical applications to the analysis of electroweak processes.

An outline of this paper is as follows:  In Section 2, I will discuss
in more detail the need for new
interactions beyond the Standard Model.   In Section 3, I will review
the principles of SMEFT that we will need for our applications.  In
Section 4,  I will describe the use of SMEFT to describe possible
quark and lepton compositeness.   In Section 5, I will describe the
application of SMEFT to the analysis of corrections to precision $Z$
physics measurements.  In Section~6, I will pause to briefly review the
possible effects of new physics on Higgs boson couplings. 
 In Section 7, I will discuss the measurement of these couplings
 through
 the application
of SMEFT to the analysis of Higgs boson processes at $\ee$ colliders.
In Section 8, I will discuss the prospects for precision Higgs boson
measurements at next-generation $\ee$ colliders.  Section 8 will
give some conclusions.

\section{The Standard Model is not complete}
\label{sec:notcomp}

To begin, I should discuss at greater depth the idea that the Standard
Model is incomplete.   Though many anomalies are discussed, there is
at 
this time no convincing evidence of a deviation from the predictions
of the Standard Model in elementary particle reactions.   The
deficiencies of the Standard Model are conceptual.   Of these, the
clearest difficulties are the facts that the Standard Model has no
explanation for the dark matter of the universe, or for the observed
preponderance of matter over antimatter.   However, the Standard Model
presents many more challenges to our  understanding.   For example,
precisely because the Higgs-fermion Yukawa couplings are
renormalizable couplings, the quark and lepton masses and mixings are
inputs to the Standard Model and cannot be explained within that
framework.

Most importantly for me, the Standard Model is incapable of explaining
the phase transition to an ordered vacuum
state that breaks the $SU(2)\times U(1)$ gauge symmetry.    The full
explanation
for this phase transition within the Standard Model is
\begin{enumerate}
  \item The most general renormalizable potential for the Higgs field
    is
    \beq
    V = \mu^2 |\Phi|^2 + \lambda |\Phi|^4 \ .
    \eeqn
  \item  The parameter $\mu^2$ satisfies   $\mu^2 < 0$.
    \end{enumerate}
The value of $\mu^2$ receives large (divergent) additive radiative corrections
with both signs.   So it is very difficult to give a coherent
explanation for the sign of $\mu^2$.   Sophisticated theorists call this
the ``gauge hierarchy problem''.  I prefer to state the problem as the fact
that we have no idea where the value of $\mu^2$ comes from.

It is not like this elsewhere in physics.   Condensed matter physics
has many examples of order-disorder phase transitions---in magnets,
superconductors and superfluids, binary alloys, liquid crystals, and
other systems.  In all cases, there is a nontrivial and fascinating
explanation for the ordering in the
ground state.   Superconductivity provides an especially interesting
example.
In 1950, Landau and Ginzburg put forward a phenomenological theory of
superconducitivity that is the model for the Higgs sector of the
Standard Model~\cite{LG}.   This is an extremely powerful and
successful theory.  It explains the thermodynamics of the phase
transition, the presence of a critical magnetic field, the distinction between
Type I and Type II superconductors and the existence of the Abrikosov
flux state.
What is does not do is give a fundamental explanation for why
superconductivity occurs.   That insight came only 7 years later, with
the work of Bardeen, Cooper, and Schrieffer~\cite{BCS}.

In the theory of the Higgs field vacuum, we are still at the
Landau-Ginzburg stage.   To go beyond this stage, we need a theory
with new particles and interactions beyond those of the Standard
Model.
 There is no rigorous argument that the universe  contains this extension,
 but this logic offers us a remarkable opportunity to discover new, hidden
 laws of nature.     We should not ignore it.

 \section{Principles of Standard Model Effective Field Theory}
 \label{sec:SMEFTprinc}

If there are new particles and interactions at high energy, how are
these reflected in the observables of the electroweak interactions?
We would like an answer to this questions  that is systematic and that
uses as few assumptions as possible about the scenario for
physics beyond the Standard Model.

One of the unexpected results of  the search for physics beyond the Standard
Model is that no new particles have yet been discovered in the energy
range of the LHC.   Although the possibilities for lighter new
particles have not yet been exhausted, this suggests that we assume
that new particles have masses above a mass scale $M$, where $M \gg
m_h$.   In this case, we can imagine integrating out the new
fields. This will leave behind
a local Lagrangian quantum field theory with the gauge symmetries of the Standard
Model and built from Standard Model fields only.  The integration out
of heavy fields may produce terms in this Lagrangian with dimension
higher than 4, corresponding to non-renormalizable interations.
Still, it is possible to compute with this Lagrangian in a
straightforward way, as long as we treat all of the coefficients  of
renormalizable and nonrenormalizable operators as free parameters to
be determined from experiment~\cite{Brivio}.   The foundations of this approach were
set out in classic papers of Ken Wilson~\cite{Wilson} and Steven
Weinberg~\cite{Weinberg}.   In the early 1980's Gasser and Leutwyler
demonstrated the power of this approach by working out in detail the
application to the low-energy scattering of $\pi$ and $K$ mesons~\cite{GasserL}.

Without any further
assumptions except that $M$ is sufficiently large, we can now
construct a general theory of new physics effects on Standard Model
precision calculations.  Again, we consider the most general
Lagrangian with $SU(3)\times SU(2)\times U(1)$ gauge invariance built
from Standard
Model fields. Consider first the part of the Lagrangian containing
operators of dimension 4 and below.  In fact, this part is nothing
more than the Standard Model itself.  The Standard Model is in fact
the most
general renormalizable Lagrangian consistent with $SU(3)\times
SU(2)\times U(1)$
gauge invariance~\cite{WeinbergSM,Nanopoulos}.   When heavy particles
are integrated out, the couplings in the renormalizable part of the
Lagrangian are shifted.  However, these shifts are unobservable,
since in any event the Standard Model couplings are fit to experiment.

Integrating out heavy particles will also generate new terms
proportional to higher dimension operators, of dimension 6, 8,
$\ldots$.
Operators of odd dimension lead to baryon- or lepton-number;
for example, these give neutrino mass terms.  I will ignore these in
the rest of this article.  Higher-dimension  operators come with dimensionful
coefficients, whose size is set by the mass scale $M$ that is
integrated out.   Then, the form of the effective Lagrangian is
\beq
\L   =   \L_{SM} +  \sum_i { {\bar c}_i\over M^2} {\cal O}_i +
\sum_j  {{\bar d}_j\over M^4} {\cal O}_j + \cdots\ ,
\eeq{effL}
where ${\bar c}_i$, ${\bar d}_j$, etc., are dimensionless coefficients.
In processes at center of mass energy $\sqrt{s}$, the
higher-dimension operators lead to effects of order $s/M^2$,
$(s/M^2)^2$, and so on.  For $M \gg \sqrt{s}$, this is a systematic
approximation scheme.
The Lagrangian \leqn{effL} defines the Standard Model Effective Field
Theory (SMEFT).

To guide intuition, consider the case of $M = 1$~TeV and ${\bar c}_i$,
${\bar d}_j$ of order 1.  Under these assumptions, the dimension-6 operators give
few-percent corrections to the Standard Model, and the dimension-8
operators give corrections of order $10^{-4}$. Then,  in practical
applications, we can ignore the operators of dimension 8 and higher
and concentrate on the effects of operators of dimension 6.    The
most general models of new physics, subject to the requirement of
large $M$, are described by a finite set of parameters $\{ {\bar
  c}_i\}$.   These assumptions can be excessively strong in some models, but it
is difficult for the ${\bar c}_i$ to take much larger values
without violating unitarity~\cite{Restimate}.

What if $M$ is not much larger than $m_h$?   In that case, the
approximation that I have described cannot be completely systematic, but
there are examples in which it is qualitatively, and even
quantitatively, correct.  We will see one example in Section~\ref{sec:PEW}.
 
In analyses that involve a large number of higher-dimension  operators,
especially when the physical significance of the scale $M$ is not
clarified by relation to a model, it is useful to set the scale of the
operator coefficients using the Higgs field vacuum expectation value
$v$.   Then I will write the SMEFT effective Lagrangian
\beq
\L   =   \L_{SM} +  \sum_i {c_i\over v^2} {\cal O}_i +
\sum_j  {d_j\over v^4} {\cal O}_j + \cdots\ .
\eeq{effLv}
In the intuitive picture suggested above, the $c_i$ will be small
parameters, of order 1\%, the $d_j$ of order $10^{-4}$, and so on. I
will use this notation in my  discussion beginning with
Sec.~\ref{sec:PEW}. 

\section{SMEFT description of lepton and quark compositeness}
\label{sec:composite}

The approximation scheme suggested by SMEFT is powerful, but it has a
difficulty.  The number of gauge-invariant dimension-6 operators is
large, and this number increases rapidly with the number of generations
and with the operator dimension.   For 1 generation, the number of independent
baryon- and lepton-number conserving operators is 59~\cite{Warsaw};
for 3 generartions, it is 2499~\cite{Alonso}.    The SMEFT approximation scheme is only
useful if there is a subset of operators that can be argued to give a
complete description of a particular problem.   The earliest examples
of the use of SMEFT to parametrize new physics are all of this type.

%%%%%%%%%%%%%%%%%%%%%%%%%%%%%%%%%%%%%%%%%%%%%%%%%%%%%%%%%%%%%%%%%%%%%%%%%
\begin{figure}
\begin{center}
\includegraphics[width=0.50\hsize]{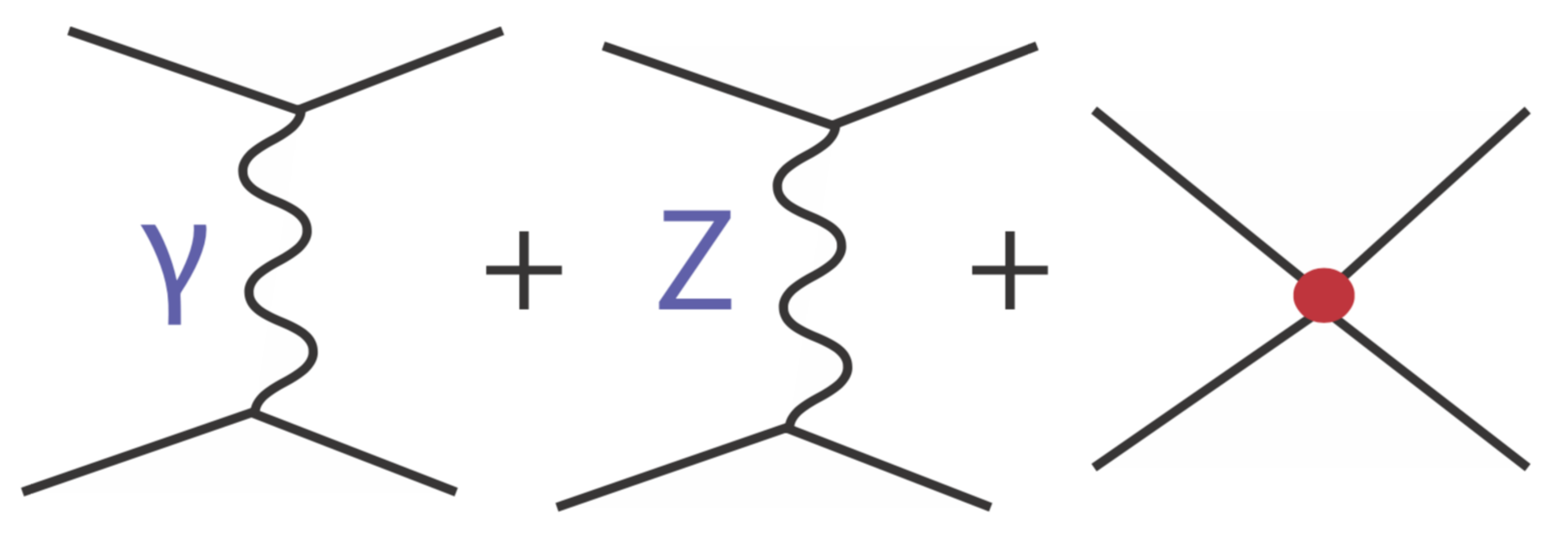}
\end{center}
\caption{Contributions to $\ee\to f\bar f$ from $s$-channel photon and
$Z$ exchange, and from a strong interaction of fermion constituents.}
\label{fig:contact}
\end{figure}
%%%%%%%%%%%%%%%%%%%%%%%%%%%%%%%%%%%%%%%%%%%%%%%%%%%%%%%%%%%%%%%%%%%%%%%%%%%

The first example came as the answer to a question posed at Snowmass
1982:  In models in which the quarks and leptons are composite, how should
one parametrize the size of composite fermion?  Up to that time,
compositeness was usually considered as modifying pointlike
electroweak couplings by the addition of form factors.   In a
leading-order scattering process
whose Standard Model amplitude would be of order $\alpha$, the
compositeness effect would be of order $\alpha \cdot s/M^2$, where $s$ is
the CM energy and $M$ is the inverse of the bound state size.
However, in addtion to weak gauge interactions, composite states bound
by a new strong interaction would also have contact interactions
involving the exchange of the bound constituents.  This strong
interaction effect would
be of order $s/M^2$, with no weak gauge suppression; see
Fig.~\ref{fig:contact}.
This observation
was described in the Snowmass proceedings~\cite{Abolins} and, more
formally, in an article by Eichten, Lane, and me~\cite{ELP}.

Assuming helicity conservation at short distances, to forbid the
generation of large masses for the light fermions, this contact
interaction is parametrized by dimension-6 current-current operators.
We noted that, for the process of Bhabha scattering, there are exactly
3 such operators, so that the process can be described by the
effective Lagrangian
\beqa
      \L &=& \L_{SM} + {2\pi\over \Lambda^2}\biggl[  \eta_{LL}\ \bar e_L
    \gamma^\mu e_L \ \bar e_L \gamma_\mu e_L + \eta_{RR}\    \bar e_R
     \gamma^\mu e_R \ \bar e_R\gamma_\mu e_R \CR
     & & \hskip 1.2in   +  2\eta_{LR} \ \bar e_L
     \gamma^\mu e_L \ \bar e_R \gamma_\mu e _R\biggr] \ ,
     \eeqa{ELPops}
  where $\Lambda$ is interpreted as the scale of compositeness and
  the operator coefficients $\eta_{IJ}$ are expected to be of order 1.
Because the Standard Model amplitude for Bhabha scattering violates
parity, the three operator coefficients can be
determined independently by fitting to the Bhabha scattering angular
distribution.   Analysis of the LEP~2 data gives 95\% CL limits on the
$\Lambda$ parameter ranging from 6 to 16~TeV depending on the choice
of nonzero values of the $\eta_{IJ}$~\cite{Bourilkov}.  In the worst case, this
corresponds to a limit on the size of the electron of
\beq
r_e <    3\times 10^{-18} \  \mbox{cm}  \  .
\eeq{esize}

In the case of quark-quark scattering, such a model-independent
analysis is not possible.   There are 17 possible contact
interactions, and these depend on the flavors and chiralities of the
interacting species.   Typically, limits are quoted under the
assumption that one operator, a universal left-handed contact
interaction, 
\beq
   \Delta\L =  \pm  {2\pi\over \Lambda} \sum_{f,f'} \bar q_{Lf}
    \gamma^\mu q_{Lf} \ \bar q_{Lf'} \gamma_\mu q_{Lf'} 
    \eeq{qleft}
   is added to the Standard Model.   However, these model-dependent
   limits are very impressive.  Using data at 13~TeV, the ATLAS
   and CMS experiments have set 95\% CL limits on $\Lambda$
   of 13 and 22~TeV for the two possible signs of the contact
   term~\cite{ATLAScontact,CMScontact}.

\section{SMEFT description of corrections to precision electroweak\\ 
  observables}
\label{sec:PEW}

Another situation in which analysis with a reduced set of dimension-6
operators makes sense comes in the study of new physics corrections
to precision electroweak interactions.   The most general effects of
new physics bring in a large number of dimension-6 operators.
However, there is a specific interesting circumstance which is
described by adding only two new operators to the Standard Model.

%%%%%%%%%%%%%%%%%%%%%%%%%%%%%%%%%%%%%%%%%%%%%%%%%%%%%%%%%%%%%%%%%%%%%%%%%
\begin{figure}
\begin{center}
\includegraphics[width=0.30\hsize]{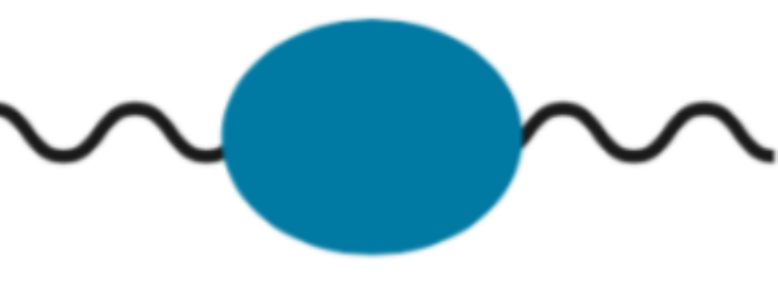}
\end{center}
\caption{A vacuum polarization diagram contributing a new physics
  correction to precision electroweak observables.}
\label{fig:vac}
\end{figure}
%%%%%%%%%%%%%%%%%%%%%%%%%%%%%%%%%%%%%%%%%%%%%%%%%%%%%%%%%%%%%%%%%%%%%%%%%%%

In many  models of new physics, the new particles couple directly to
the Higgs sector but with very small couplings to light quarks
and leptons.   This applies, for example, to models of an extended or
composite Higgs sector, and models with new quarks, leptons, or
vectorlike fermions.   In the limit in which these light quark
couplings can be ignored, the new physics
corrections to electroweak reactions at low energies and at the $Z$ pole
come only from  vacuum polarization diagrams, as in
Fig.~\ref{fig:vac}.   Lynn, Stuart, and I  described  this
limit by labelling these diagrams  as ``oblique corrections''~\cite{LPS}.   This
terminology  calls attention to a simple but important class of new physics
effects that are amenable to general analysis. 

%%%%%%%%%%%%%%%%%%%%%%%%%%%%%%%%%%%%%%%%%%%%%%%%%%%%%%%%%%%%%%%%%%%%%%%%%
\begin{figure}
\begin{center}
\includegraphics[width=0.95\hsize]{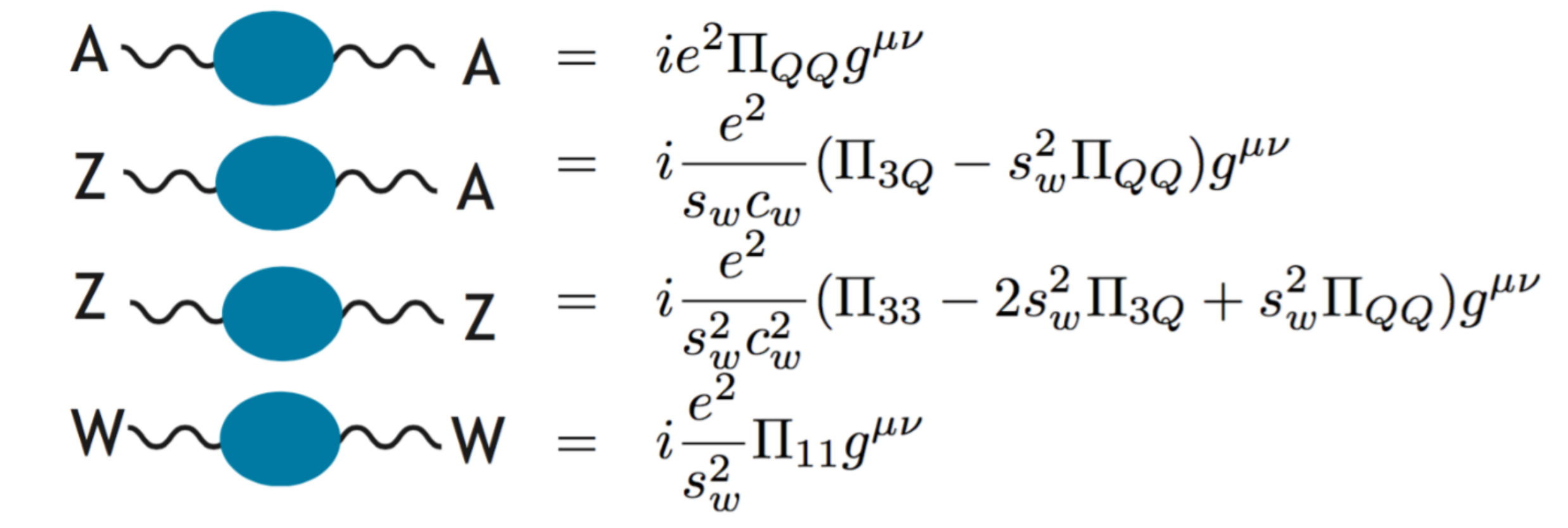}
\end{center}
\caption{The photon, $Z$, and $W$
  vacuum polarizations decomposed  in terms of their $SU(2)\times U(1)$
  components. The notation on the right-hand side is explained further
  in the text.}
\label{fig:vpols}
\end{figure}
%%%%%%%%%%%%%%%%%%%%%%%%%%%%%%%%%%%%%%%%%%%%%%%%%%%%%%%%%%%%%%%%%%%%%%%%%%%

In fact, that analysis turns out to be very straighforward. As
Takeuchi and I pointed out~\cite{PT}, it does
not require any sophisticated operator counting, but only a glance at the
Taylor expansion of the vacuum polarization amplitudes in powers of
$q^2$.   There are four relevant vacuum polarization amplitudes;
define these according to their $SU(2)$ quantum numbers as shown in
Fig.~\ref{fig:vpols},  In the figure, 
$s_w = \sin\theta_w$, $c_w = \cos\theta_w$. I have omitted
terms proportional to $q^\mu q^\nu$ that, in any event, give zero when
contracted with light fermion lines.   The subscripts
1
and 3 refer to currents of the gauge $SU(2)$ symmetry and $Q$ refers
to the electric charge current.   

 If we take $M$ to be the scale of
new particle masses,  the four amplitudes have Taylor expansions in
$q^2/M^2$  of
the form
\beqa
  \Pi_{QQ} &=&    A q^2 + \cdots \CR
  \Pi_{3Q} &=&   B  q^2 + \cdots \CR
  \Pi_{33} &=&    C + D  q^2 + \cdots \CR
  \Pi_{11} &=&    E + F  q^2 + \cdots \ .
  \eeqa{PiTaylors}
  The zeroth-order terms in the first two lines vanish due to electric
  current conservation.
  Of the 6 coefficidents, 3 linear combinations are fixed by the renormalization of the 3
  basic parameters of the Standard Model $g$, $g'$, and $v$.   The
  remaining 3 linear combinations will be finite in a renormalizable
  extension of the Standard Model.   These are canonically defined as 
  \beqa
     S &=& {16\pi\over m_Z^2} \biggl[\Pi_{33}(m_Z^2) - \Pi_{33}(0) -
     \Pi_{3Q}(m_Z^2) ]\biggr] \CR
    T &=& {4\pi\over s_w^2 m_Z^2} \biggl[\Pi_{11}(0) - \Pi_{33}(0) ]\biggr] \CR
    U &=& {16\pi\over m_Z^2} \biggl[\Pi_{11}(m_Z^2) - \Pi_{11}(0) -
    \Pi_{33}(m_Z^2) + \Pi_{33}(0) ]\biggr] \ . 
    \eeqa{STdefin}
    
    The parameters $S$ and $T$ have appealing physical
    interpretations.  $T$ indicates  the correction to the Standard
    Model relation $m_W = m_Z \cos\theta_w$ that reflects its
    custodial $SU(2)$ symmetry~\cite{SSVZ}.   $S$ indicates in a
    dimesionless way the size of the (custodial symmetry-invariant)
    new physics sector.

    The leading oblique corrections to electroweak observables can
    then be expressed as linear combinations of $S$, $T$, and $U$.  It
    is useful to make reference to the value of $\theta_w$ constructed
    from the best-measured observables $\alpha$, $G_F$, and $m_Z$,
    \beq
         \sin^2 2\theta_0  \equiv  {\alpha(m_Z^2)\over \sqrt{2} G_F
           m_Z^2}\ .
      \eeq{szerodef}
   Then it is possible to represent the effect of general oblique
   corrections as deviations from the values of observables predicted
   by the Standard Model with this standardized value of $\sstw$. For example,
    \beqa
         {m_W^2\over m_Z^2}  - c_0^2 &=&  {\alpha  c_w^2\over c_w^2 -
           s_w^2}\bigl(
               - \half S + c_w^2 T + {c_w^2 - s_w^2\over 4 s_w^2} U \bigr) \CR
    s_*^2 - s_0^2 &=&  {\alpha  \over c_w^2 -
           s_w^2}\bigl(
         {1\over 4}  S - s_w^2 c_w^2 T \bigr) \ ,
         \eeqa{STobs}
    where $s_*^2$ is the value of $\sstw$ extracted from the $Z$
    resonance 
    polarization asymmetries.   By fitting deviations from the
    Standard Model predictions to these formulae, we can put
    constraints on the full set of models to which the assumptions of
    the oblique approximation apply. 

    In SMEFT, the parameters $S$ and $T$ are represented by adding two
    dimension-6 operators to the Standard Model Lagrangian,
    \beq
    \L = \L_{SM} + {c_T\over 2v^2}(\Phi^\dagger  \Dlr^\mu \Phi)
    (\Phi^\dagger  \Dlr_\mu \Phi) +  {16 s_w\over c_w} {c_{WB} \over
      v^2}
    \Phi^\dagger  t^a \Phi \, W^a_{\mu\nu} B^{\mu\nu}  \ .
    \eeq{STLag}
    Here, the notation is as in \leqn{effLv},
    $c_T$ and $c_{WB}$ are dimensionless operator coefficients, $\Phi$
    is the Standard Model Higgs doublet field, $W^a_{\mu\nu}$ and
    $B_{\mu\nu}$ are the $SU(2)$ and $U(1)$ field strengths, and
    \beq
    \Phi^\dagger  \Dlr_\mu \Phi =  \Phi^\dagger  \D_\mu \Phi
    -  D_\mu \Phi^\dagger \ \Phi \ .
    \eeq{Dlrdef}
    I will clarify the relation of this truncated Lagrangian to the
    full SMEFT Lagrangian in Sec.~\ref{sec:Higgs}.   The relation
    between the SMEFT coefficients and the $S$ and $T$ parameters is
    \beq
    \alpha \, S = 4 s_w^2 (8 c_{WB})   \qquad
    \alpha\, T = c_T \  .
    \eeq{STarecs}

    The parameter $U$ is doubly suppressed,
    requiring both direct effects of heavy new particles and custodial
    $SU(2)$ violation.  In the SMEFT context, $U$ turns out to be
    generated by a dimension-8 operator.   When $U$ is included in
    fits to electroweak data in any event, its value is consistent
    with 0. 

  Some guidance about the expected sizes of $S$ and $T$ is given by
    the expressions for these quantities in specific models.   For
    example, for one new heavy  electroweak doublet $(N,E)$,
    \beq
          S = {1\over 6 \pi} \qquad     T = {|m_N^2 - m_E^2|\over
            m_Z^2}  \ .
          \eeq{STforNE}
          
       Since the top quark and the Higgs boson have only tiny direct
       couplings to the light generations, we can express the
       contribution to electroweak observables from these Standard
       Model particles in the $S$, $T$ framework.  For the top quark,
\beq
S = {1\over 6\pi} \log {m_t^2 \over m_Z^2}  \qquad
T = {3\over  16\pi s_w^2 c_w^2}  {m_t^2\over m_Z^2} \ ;
\eeq{topST}
for the Higgs boson
\beq 
S = {1\over 12\pi} \log {m_h^2 \over m_Z^2}  \qquad
T = - {3\over  16\pi c_w^2}  {m_h^2\over m_Z^2} \ .
\eeq{HiggsST}
Although the top quark and Higgs boson masses are by no means much
larger than $v$, as would be needed to give  formal justification to the SMEFT
approximation,
these formulae turn out to be a very good
representation of the effects of the top quark and the Higgs boson on
the precision electroweak fit.

%%%%%%%%%%%%%%%%%%%%%%%%%%%%%%%%%%%%%%%%%%%%%%%%%%%%%%%%%%%%%%%%%%%%%%%%%
\begin{figure}
\begin{center}
\includegraphics[width=0.99\hsize]{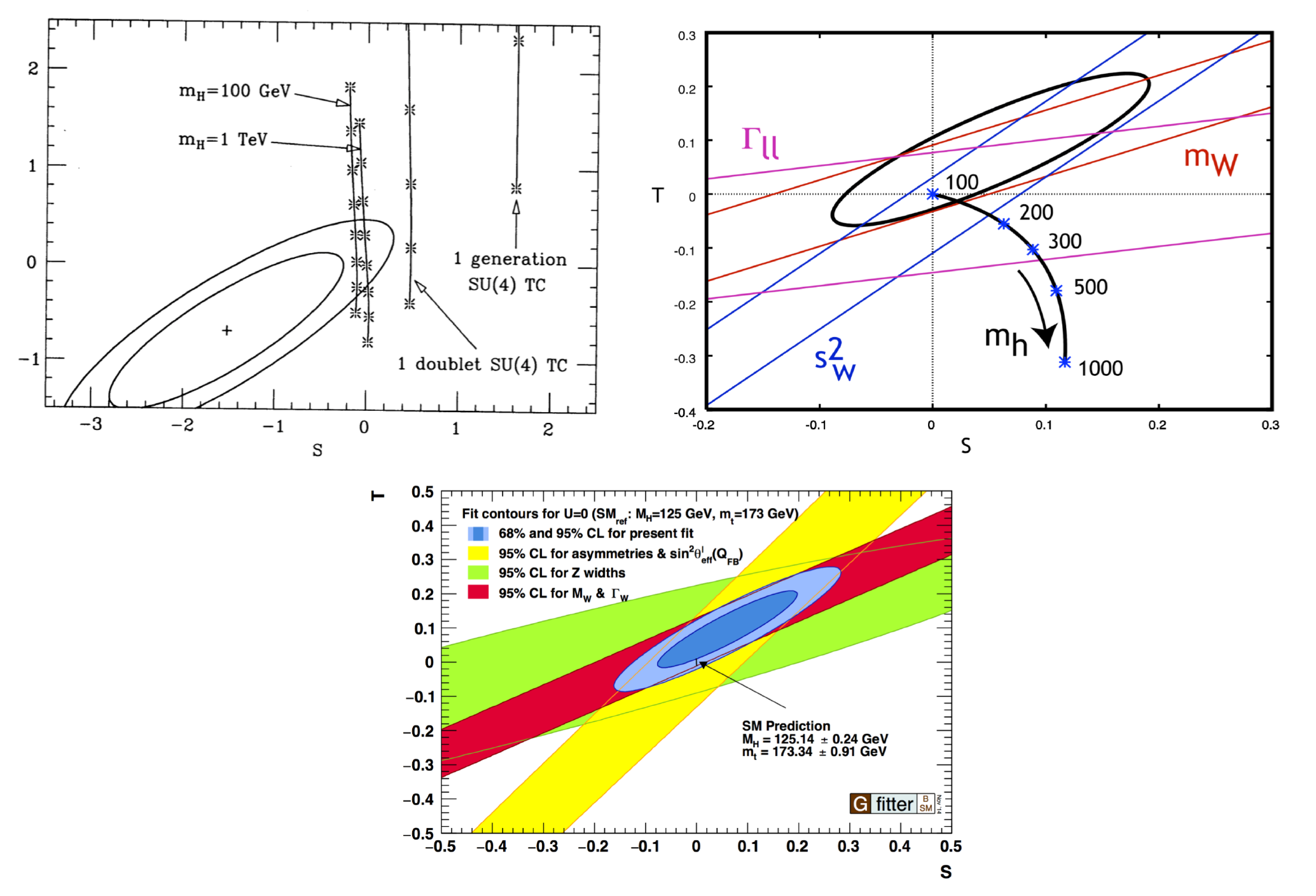}
\end{center}
\caption{ST fits to the precision electroweak data from
  1991~\cite{PT}, 2008 (based on the results of \cite{LEPEWZ}) 
  and 2014~\cite{Gfitter}. Note the changes of scale and the changing 
interpretation  as new inputs are added.}
\label{fig:STfits}
\end{figure}
%%%%%%%%%%%%%%%%%%%%%%%%%%%%%%%%%%%%%%%%%%%%%%%%%%%%%%%%%%%%%%%%%%%%%%%%%%%

The progress of the $S$, $T$ fit to electroweak observables
since the days of the earliest LEP
and SLC data indicates clearly the impressive growth of our knowledge.
Figure~\ref{fig:STfits} shows the 1991 $S$, $T$ fit from \cite{PT},  a
2008 fit based on the final results of the $Z$ resonance parameters
reported in \cite{LEPEWZ}, and a 2014 fit by the Gfitter
Collaboration~\cite{Gfitter}.
Note the changes in scale in the three plots.   The first plot
predicts a top quark mass in the range  120--180~GeV.  The second plot
uses the by-then measured value of the top quark mass and predicts a
value of the Higgs boson mass below 140~GeV.   The third plot gives
the current status of the electroweak fit, in excellent agreement with
the Standard Model in accord with the known values of the top quark
and Higgs boson masses. 

\section{The Higgs boson as a probe of physics beyond
  the Standard Model}

The next logical target for electroweak precision measurement is the
Higgs boson.   It is widely appreciated that the verification of the
Standard Model will not be complete without a detailed study of the
properties of the Higgs boson.    I feel that it is less well
appreciated that precision measurement of the Higgs boson couplings
gives a remarkable opportunity for the discovery of new physics,
beyond the capabilities of the LHC experiments.  In this section, I
will explain my viewpoint on this question.

First of all, the discovery of the Higgs boson at the LHC in
2012~\cite{ATLASh,CMSh} puts us in a new situation with respect to the
Standard Model.  The complete set of particles predicted by the
Standard Model have been discovered, and their masses have been
measured accurately.  In particular, the mass of the Higgs boson has
been determined as \cite{ATLAShm,CMShm,PDGHiggs}
\beq
    m_h = 125.10 \pm  0.14\ \mbox{GeV} \  . 
    \eeq{Higgsmass}
With this measurement, all of the parameters of the Standard Model are
specified to part-per-mil accuracy.  From these parameters, we can
predict the properties of the Higgs boson in detail without
ambiguity.  Any deviation from these predictions would be a signal of
new interactions beyond the Standard Model.

The fact that the
Higgs boson mass is close to 125~GeV has the
consequence, according to the Standard Model, that this particle has 10
distinct decay modes with branching ratios greater than $10^{-4}$.
Five of these modes, the decays to $ZZ^*$, $WW^*$, $b\bar b$,
$\tau^+\tau^-$, and $\gamma\gamma$, and the Higgs boson couplings to
$gg$ and $t\bar t$, have already been observed at the LHC.   The
couplings are consistent with the Standard Model predictions up to
uncertainties of 10-30\%~\cite{CMScomb,ATLAScomb}.

  This is a very impressive increase in our knowledge, but it cannot
  be taken as evidence against physics beyond the Standard Model.
  In Sec.~\ref{sec:SMEFTprinc}, we saw that observable effects of new
  physics on the observed Higgs boson are associated with dimension-6 SMEFT
  operators and are generically of a few percent in size.   So, the
  current level of agreement with the predictions of the Standard
  Model is just that expected in any extension of the Higgs sector.
  To use the Higgs boson to probe for new physics, we need to push the
  measurement uncertainties down below  the 1\% level.

%%%%%%%%%%%%%%%%%%%%%%%%%%%%%%%%%%%%%%%%%%%%%%%%%%%%%%%%%%%%%%%%%%%%%%%%%
\begin{figure}
\begin{center}
\includegraphics[width=0.60\hsize]{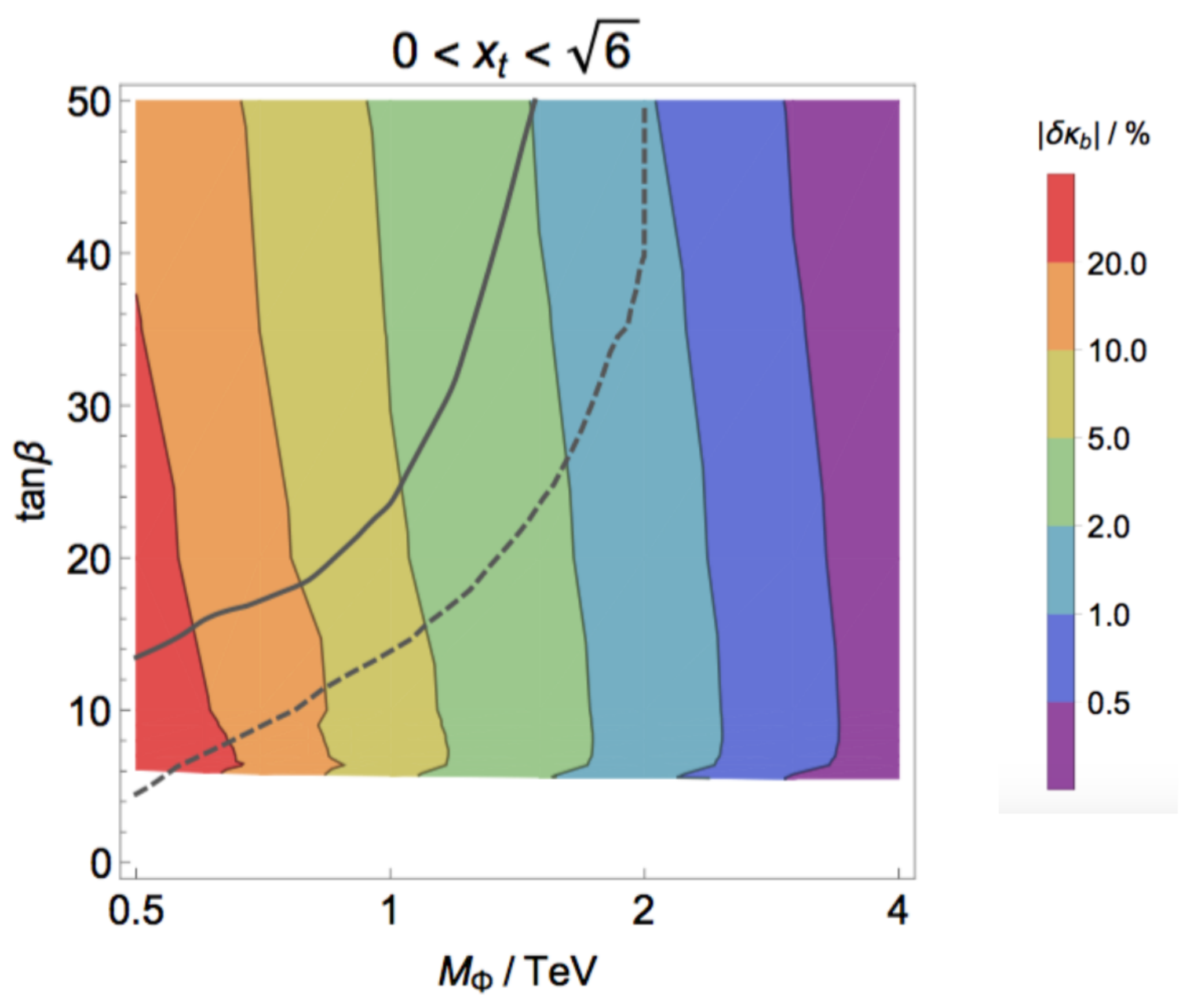}
\end{center}
\caption{Fractional deviations of the $hb\bar b$ coupling, in \%, from the Standard Model
  expectation in a class of supersymmetric models studied by Wells and
 Zhang~\cite{WellsZhang}, as a function of model parameters.  The
 models in the upper left, above the solid line, are excluded by LHC
 searches.  The dotted line shows the exclusion contour expected from  the HL-LHC.}
\label{fig:Wells}
\end{figure}
%%%%%%%%%%%%%%%%%%%%%%%%%%%%%%%%%%%%%%%%%%%%%%%%%%%%%%%%%%%%%%%%%%%%%%%%%%%

%%%%%%%%%%%%%%%%%%%%%%%%%%%%%%%%%%%%%%%%%%%%%%%%%%%%%%%%%%%%%%%%%%%%%%%%%
\begin{figure}
\begin{center}
\includegraphics[width=0.44\hsize]{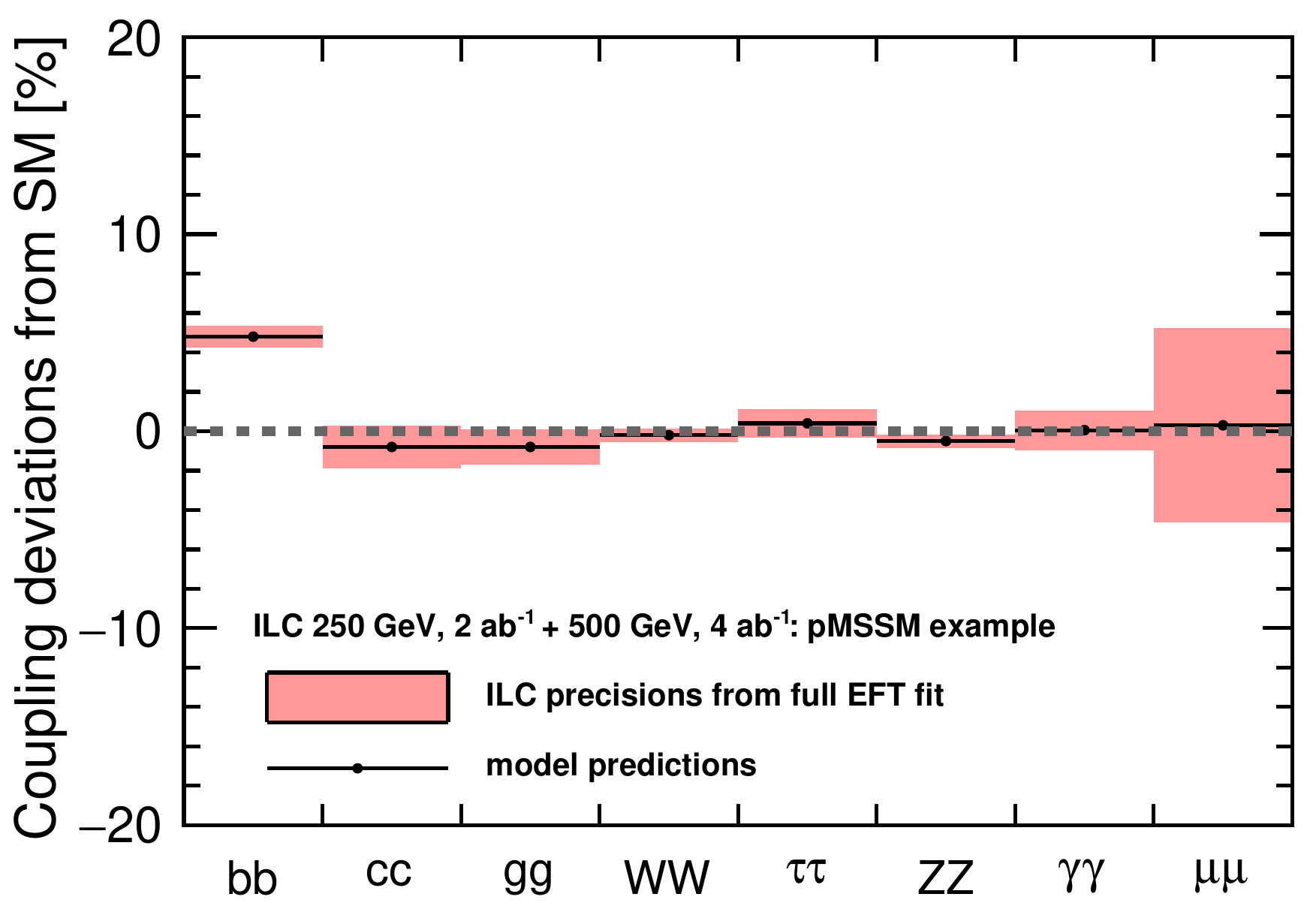} \ 
\includegraphics[width=0.44\hsize]{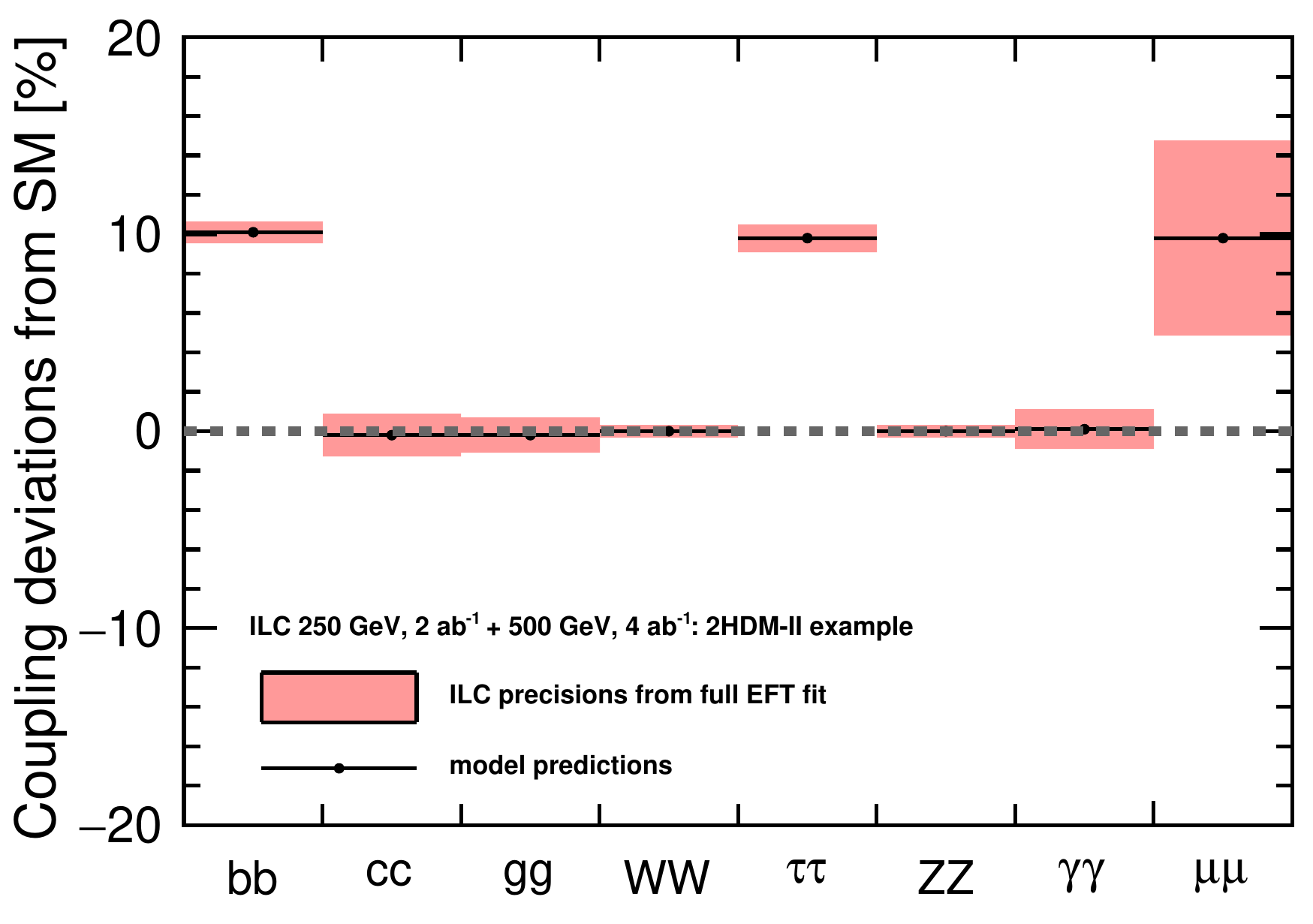} \CR
\includegraphics[width=0.44\hsize]{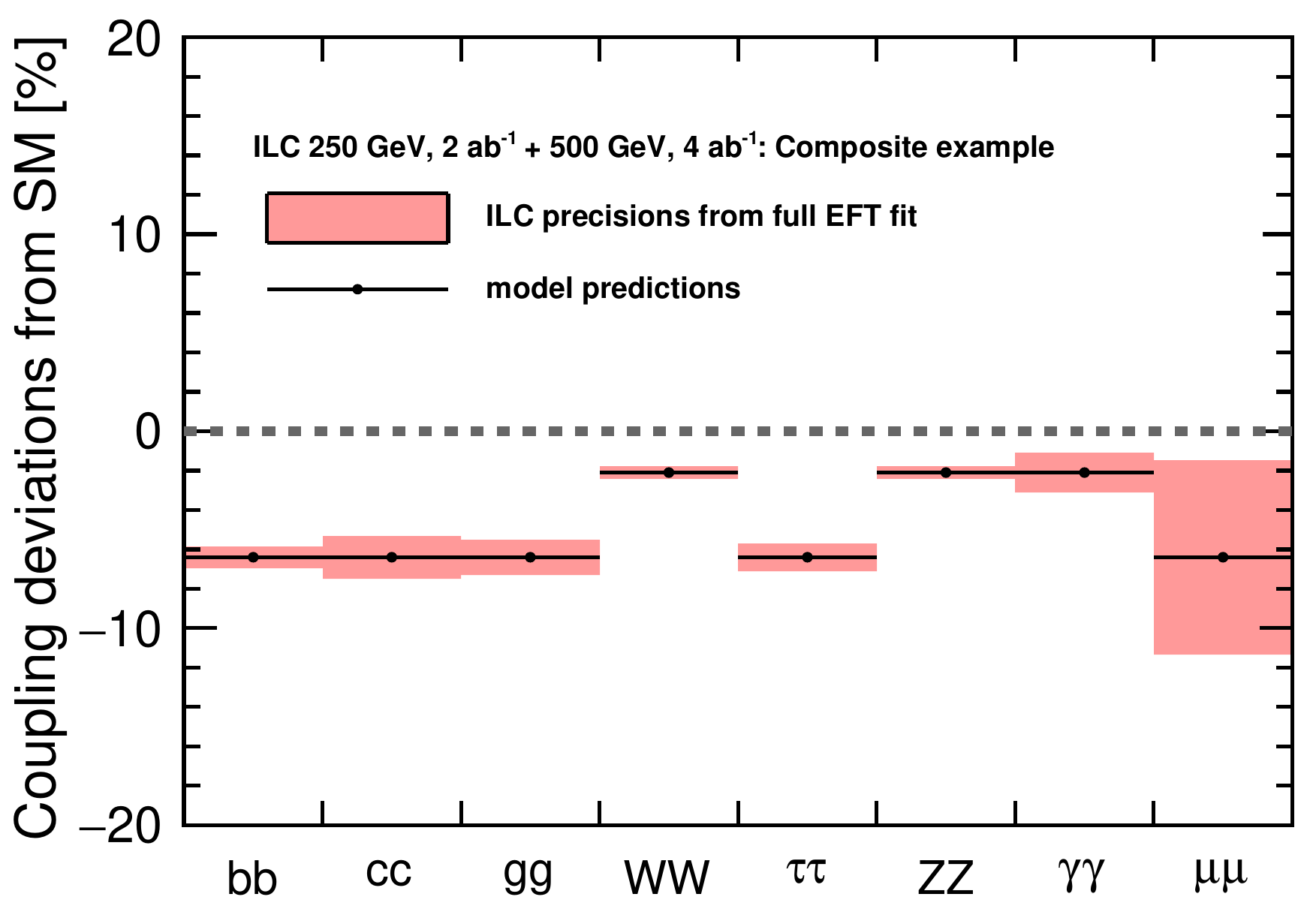} \ 
\includegraphics[width=0.44\hsize]{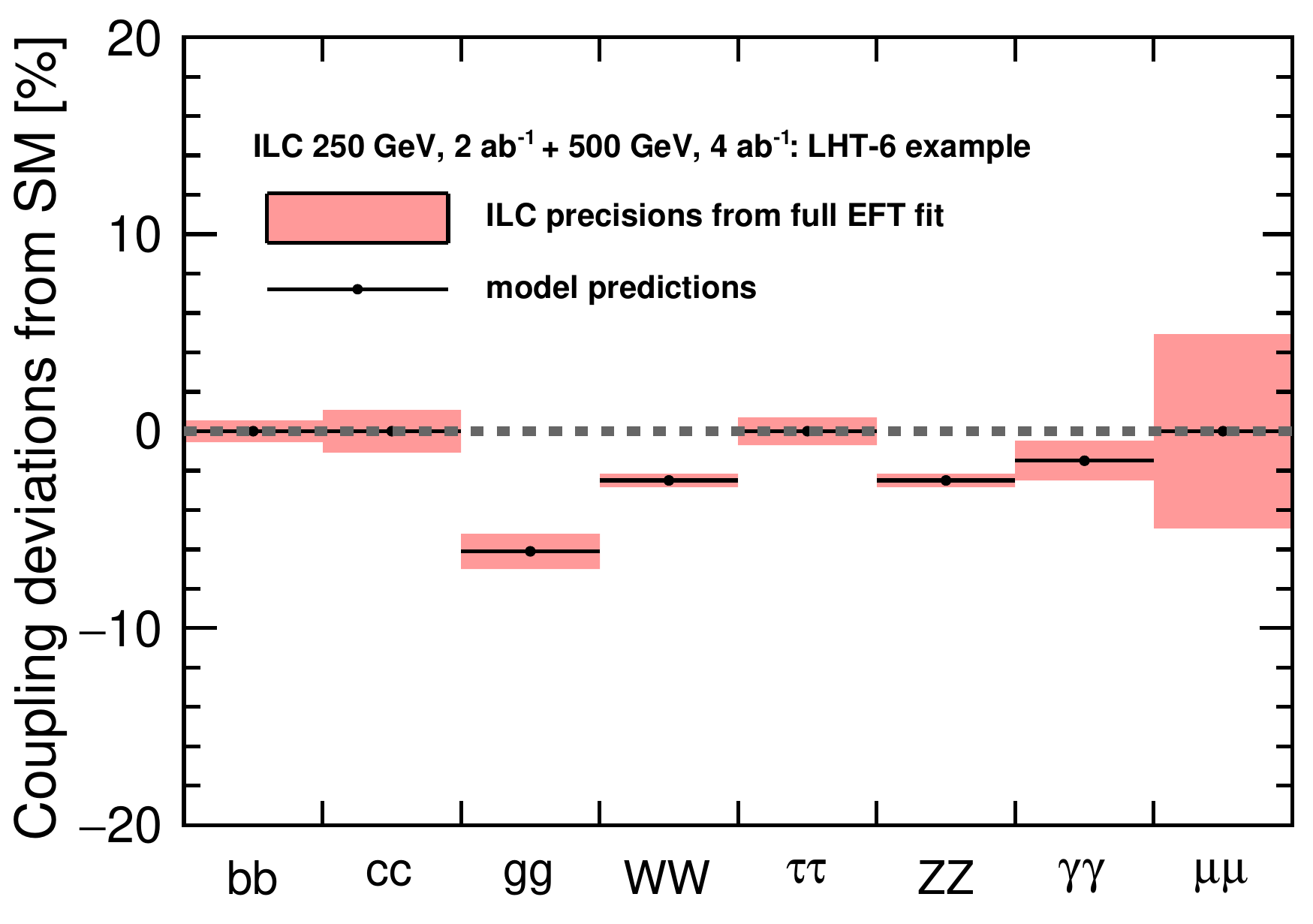}
\end{center}
\caption{Fractional deviations of 8 Higgs boson couplings, in \%, from
  the Standard Model expectations, in a variety of new physics
  models.  The four models shown are representative
  supersymmetric models, two-Higgs doublet models, composite Higgs
  models, and Little Higgs models.  In all cases, the parameters are
  chosen so that the new particles predicted by these models are not
  expected to be discovered at the HL-LHC.}
\label{fig:Higgscases}
\end{figure}
%%%%%%%%%%%%%%%%%%%%%%%%%%%%%%%%%%%%%%%%%%%%%%%%%%%%%%%%%%%%%%%%%%%%%%%%%%%

  However, if we can perform Higgs measurements sufficiently precisely
  to meet this criterion, considerable insight is available.   There
  are two important points to be made here.  First,  the study of new
  physics effects on the Higgs boson couplings  gives a window on  new
  physics
  that is different from the search for new particles.  Though it is
  tempting to compare the ``reach'' of direct searches and precision
  measurements, this is too simplistic a view.  The point is
  illustrated in Fig.~\ref{fig:Wells}, from \cite{WellsZhang}.   The
  colored bands show the expected variation of the $hb\bar b$ coupling
  from the Standard Model prediction in a class of supersymmetric
  models with $b$-$\tau$ Yukawa unification.  The region in the upper
  left-hand corner, bounded by the solid line, is the part of the
  parameter space excluded by the LHC experiments in Run 2.   At the
  end of the HL-LHC running, the region down to the dotted line is
  expected to be explored.  Below these curves, though,  there is a whole space of models
  in which the $b$
squarks have multi-TeV masses and cannot be discovered by LHC
searches but, at the same time, they produce modifications of
the  $hb\bar b$ coupling of 1-3\%  that are potentially observable.

The second point is that 
  different extensions of the Higgs sector
  have their most important effects on different Higgs boson decay
  modes.   New physics models have parameter freedom, and the effects
  on the Higgs couplings vary over the parameter space.  But, in
  general,
  \begin{itemize}
    \item Higgs couplings to $b$, $\tau$ are modified by
      supersymmetric and 2-Higgs-doublet models
      \item Higgs couplings to $W$, $Z$ are modified by composite
        Higgs models and mixing with scalar singlets
     \item Higgs couplings to $g$, $\gamma$, $t$ are modified by  top
       quark partners and symmetry-breaking models based on top
       condensation
     \end{itemize}
  This subject is reviewed in more detail in  \cite{myCERNSS}.
The point  is illustrated in Fig.~\ref{fig:Higgscases}, from \cite{Barklow1},
 which shows the pattern of deviations of the Higgs couplings from the
 Standard Model predictions in four specific new physics models.
 In all four cases, the new particles associated with the model
 are expected to be out of the
 reach of the HL-LHC.    The error intervals shown are those expected
 from measurements at the International Linear Collider, to be
 discussed in Sec.~\ref{sec:ILC}.   With sufficient, achievable,
 precision, the study of Higgs boson couplings can not only
 demonstrate that the Standard Model is modified but also can give us guidance on
 the type of model that solves the conceptual problems of the Higgs
 theory discussed in Sec.~\ref{sec:notcomp}.

\section{SMEFT analysis of Higgs boson reactions at $\ee$ colliders}
\label{sec:Higgs}

I will now discuss how to use SMEFT to extract the values of the Higgs
boson couplings from observables measured at colliders.  The Higgs
couplings cannot be read off directly from measurements because some
needed information is missing. 
In particular, the total width of the Higgs
boson is expected in the Standard Model to be about 4.3~MeV, a value
too small to be measured directly from the width of the resonance
observed in collider detectors.   To extract the total width of the
Higgs boson, which provides the normalization of all partial widths,
we need a framework in which to fit the various measurements.   In the
best case, this framework would not make strong assumptions about the
nature of new physics that generates corrections to the predictions of
the Standard Model.    I will now present the use of SMEFT to provide that
framework.

I will concentrate here on  the extraction of Higgs boson couplings
from precise measurements of the Higgs boson at next-generation $\ee$ colliders.
The extraction of Higgs couplings from LHC data is discussed in
\cite{HLLHCproj}, together with projections for the results expected
from experiments at the HL-LHC.  Those experiments will greatly
improve our knowledge, but their interpretation will be
model-dependent, and they are expected to reach only the few-\%
level of
uncertainty, insufficient to demonstrate the existence of new physics
corrections
at the size expected from the examples of the previous section. 
  It is the future $\ee$ experiments that will really
have the power to challenge the Standard Model.

At this time, there are four proposals for $\ee$ ``Higgs factories''
under serious consideration at different sites around the world.   Two
of these are circular $\ee$ colliders of roughly 100~km circumference,
CEPC in China~\cite{CEPC1,CEPC2} and FCC-ee~\cite{FCCee} at CERN.  The other
two are linear $\ee$ colliders, ILC in Japan~\cite{ILC,ILCprec} and CLIC
at CERN~\cite{CLIC}.   The technical implementation differs
among the 4 proposals, but all have similar goals, including the
high-precision study of the reaction $\ee\to Zh$. 
The peak of the cross section for the process $\ee\to Zh$ is at a
center of mass energy 250~GeV.   Thus, a 250~GeV $\ee$ collider, well
within the capabilities of current technologies, can produce a large
sample of events in which the Higgs boson is produced together with
a $Z$ boson.

The $\ee\to Zh$ reaction is an exceptionally clean setting in which
to study the Higgs boson.   To a first approximation (with a smooth
and precisely
calculable background) any $Z$ boson observed at a lab energy of
110~GeV is recoiling against a Higgs boson.  One simply needs to
remove the $Z$ boson from the event and see what is left to measure
the quantities
\beq
\sigma(\ee\to Zh) BR(h \to A\bar A)   
\eeq{sigmaBR}
for all Higgs boson decay products $A\bar A$.   The ratios of these
quantities give the Higgs boson branching ratios.   If  we can also
determine the Higgs total width, we can find all of the partial widths
and use these to extract the Higgs boson couplings.

A simple method to find the Higgs total width $\Gamma(h)$ is to assume that each
individual Higgs coupling  $g(hA\bar A)$  is modified from its Standard Model value by
a multiplicative constant $\kappa_A$.    This parametrization has the appealing
property that the quantities
\beq
\sigma(\ee\to Zh)  \quad \mbox{and} \quad  \Gamma(h\to ZZ^*)
\eeq{affected}
are both proportional to  $\kappa_Z^2$.  The branching ratio of the
Higgs boson to $ZZ^*$ is given by 
\beq 
          BR(h\to ZZ^*) = \Gamma(h\to ZZ^*) / \Gamma(h)  \ . 
\eeqn
We can measure $\sigma(\ee\to
Zh) $ by countinng recoil $Z$ bosons, and we can measure $BR(h\to
ZZ^*)$ by identifying $Z$ bosons among the Higgs decay products.
Then
in the quantity 
\beq
{\Gamma(h\to ZZ^*)\over BR(h\to ZZ^*)} 
\eeq{findGamma}
our assumption would imply that 
the factors of $\kappa_Z^2$ cancel out and the result is directly proportional
to $\Gamma(h)$. 

There is a problem, though, that this strategy is not completely
model-independent.  In the Standard Model, the $hZZ$ coupling has the
structure $h Z_\mu Z^\mu$, but in general two independent Lorentz
structures are possible,
\beq
     \L_{hZZ} =   (1 + \eta_Z) {m_h^2\over v}\, h Z_\mu Z^\mu + \half
     \zeta_Z {1\over v} \, h Z_{\mu\nu} Z^{\mu\nu}  \ ,
     \eeq{LhZZ}
  where $Z_{\mu\nu}$ is the $Z$ field strength tensor.  There are two
  parameters, $\eta_Z$ and $\zeta_Z$, that represent possible new
  physics corrections.   Both can arise from dimension-6 SMEFT
  operators, so they are arguably on equal footing.  The $\zeta_Z$
  term leads to a  momentum-dependent vertex that gives very different
  corrections to the two quantities in \leqn{affected}.  If 
  $\zeta_Z$ is nonzero,  the dependence of the quantities in
  \leqn{affected} on new physics is not a simple overall
  multiplicative factor and the 
  argument in the previous paragraph does not go through.  What
  formalism can replace it?

  It would be attractive to use the dimension-6 SMEFT coefficients as
  the parameters in a framework for fitting the Higgs width and
  couplings.  At first sight, this seems out of reach.  I have
  explained at the beginning of Sec.~\ref{sec:composite} that the
  number of dimension-6 SMEFT coefficients is very large.  However, in
  2016, Tim Barklow proposed that, since $\ee$ annihilation processes
  are sensitive to only a subset of these operators,  and since $\ee$
  colliders allow a very large number of independent measurements, a
  fit to the relevant set of coefficients can be completely
  constrained.

  This approach was worked out in detail in
  \cite{Barklow1,Barklow2}.   We are concerned with electroweak
  processes, so it suffices to consider the new physics corrections at
  the tree level.   We make use of CP-even observables. 
  CP-violating terms contribute to these only in order $c_i^2$, and it
  is possible, by measuring CP-odd observables, to check that these
  coefficients are small enough that their effects can be ignored.
 The relevant dimension-6 operators will be those that are built from
 the fields of $h$, $W$, $Z$, $\gamma$, and $e_{L,R}$ and those that
 contribute to Higgs boson decays at tree level.  This set of
 operators can be reduced using the Standard Model equations of
 motion.   From these considerations,  we find a parameter
 set consisting of the 4
 Standard Model parameters $g$, $g'$, $v$, and
 $\lambda$, plus 18 dimension-6 operator coefficients.

 The
 dimension-6 terms involving only Higgs fields are
 \beq
 \Delta\L_1 = {c_H \over 2 v^2} \del^\mu (\Phi^\dagger \Phi) \,
 \del_\mu  (\Phi^\dagger \Phi)  +  {c_T\over 2 v^2} ( \Phi^\dagger
 \Dlr^\mu \Phi)  ( \Phi^\dagger
 \Dlr_\mu \Phi)  -  {c_6 \lambda\over v^2 } (\Phi^\dagger\Phi)^3 \ .
 \eeq{Lsixone}
 The additional terms with gauge fields can be reduced to
 \beqa
 \Delta\L_2  & = & { 4 c_{WW}\over v^2 } \Phi^\dagger \Phi W^a_{\mu\nu} W^{a
   \mu\nu} + 16 {s_w c_{WB} \over c_w v^2} \Phi^\dagger t^a  \Phi
 W^a_{\mu\nu} B^{ \mu\nu}\CR
   & & \hskip 0.2in   + {4 s_w^2 c_{WW}\over  c_w^2 v^2 } \Phi^\dagger
   \Phi
   B_{\mu\nu} B^{ \mu\nu} + {4  g c_{3W}\over v^2} W^a_{\mu\nu}
   W^{b\nu}{}_\rho W^{c\rho \mu} \ ,
   \eeqa{Lsixtwo}
The  dimension-6 terms with electrons and Higgs fields are  
  \beqa
 \Delta\L_2  & = &i { c_{HL}\over v^2 } (\Phi^\dagger \Dlr^\mu
  \Phi)\,( L^\dagger \gamma_\mu L) +
  { 4i  c^\prime_{HL}\over v^2 } (\Phi^\dagger t^a \Dlr^\mu
  \Phi)\,( L^\dagger \gamma_\mu t^a  L) \CR
 & & \hskip 0.1in + i { c_{HE}\over v^2 } (\Phi^\dagger \Dlr^\mu
  \Phi)\,(e^\dagger \gamma_\mu e) + {c_{L L_\mu}  \over v^2} (L^\dagger
  t^a \gamma^\mu L)\, (L_\mu^\dagger \gamma_\mu t^a L_\mu)
 \eeqa{Lsixthree}
 These terms allow explicit violation of the ``oblique'' assumption
 mentioned in Sec.~\ref{sec:PEW}.
 The additional terms modifying Higgs decay amplitudes are
 \beq
 \Delta L_4 =  -  {c_{\tau \Phi} y_\tau\over v^2}
 (\Phi^\dagger \Phi)\,  L_\tau^\dagger\cdot \Phi \tau_R  +
  {c_{b\Phi} y_b\over v^2}
  (\Phi^\dagger \Phi)\,  Q_b^\dagger\cdot \Phi b_R  \ ,
  \eeq{Lsixfour}
  and similar operators for $c$ and $\mu$.
There are several independent dimension-6 operators that
  contribute to the amplitude for $h\to gg$.  These are not
  distinguishable in this analysis, since only the on-shell $h\to gg$
  amplitude contributes to the observables.  We can represent this
  degree of freedom by adding
  \beq
  \delta\L_5 =   { 4c_{gg}\over v^2}  G_{\mu\nu}^a G^{a \mu\nu} \ .
  \eeq{Lsixfive}
  FInally, additional coeffiients $c_{Hf}$ similar to those in
  \leqn{Lsixthree}
multiplying dimension-6
  operators that couple the Higgs current to other flavors appear in
  the tree-level expressions for the partial widths $\Gamma(h\to
  WW^*)$ and $\Gamma(h\to ZZ^*)$.   Fortunately, only two linear
  combinations of these coefficients appear, and the same two linear
  combinations appear in the dimension-6 corrections to the $W$ and
  $Z$ total widths.
  
Of the 18 dimension-6 coefficients introduced here, $c_6$ does not appear in
single-Higgs boson observables.  Its role to shift the Higgs
self-coupling.   The parameter  $c_{L L_\mu}$ is related to one of the
the
$\Lambda$ parameters introduced in Sec.~\ref{sec:composite}.  It
is already strongly constrained by studies of the reaction
$\ee\to \mu^+\mu^-$, and that constraint is expected to become
about 100 times stronger at next-generation $\ee$ colliders~\cite{ILCprec}.  Thus, I will
ignore these two parameters here.   This leaves 4+16 parameters that
need to be determined from $\ee$ collision data.

It is difficult to find 20 independent high-precision measurements of
Higgs processes, even at $\ee$ colliders.   However, in this
formalism, the SMEFT Lagrangian is the Lagrangian for all of
electroweak physics, not only for the Higgs sector. In fact, we can
use data from all electroweak processes to constrain the 20
parameters.  The analysis is worked out in detail in \cite{Barklow1}.
From precision electroweak measurements of the $Z$ and
$W$, we have the 8 well-determined quantities
\beq
     \alpha\ , \ G_F \ , \   m_Z\ , \ m_W\ , \ A_e\ , \ \Gamma(Z\to
     \ee)\ ,\  \Gamma_W\ , \ \Gamma_Z   \ .
     \eeq{eightPEW}
Note that, of these quantities, only $G_F$, $\Gamma_W$, and $\Gamma_Z$ make
reference to any fermion other than the electron.   Thus, this
strategy makes no assumption about lepton flavor universality or
any other possible regularity of the electroweak couplings.
Measurements of the triple gauge vertices in $\ee$ constrain three
additional
parameters.  These are combinations of $c_{WB}$, $c_{WW}$ and the
$c_{HL,E}$ distinct from those that contribute to the precision
electroweak observables.  

From single-Higgs processes, the $\ee$ experiments will separately
measure the Higgs mass $m_h$, the
total cross section for $\ee\to Zh$ and the $\sigma\times
BR$ for this reaction in the $b\bar b$, $c\bar c$, $\tau^+\tau^-$,
$gg$,   $WW^*$, $ZZ^*$, $\gamma\gamma$,  $Z\gamma$,   and $\mu^+\mu^-$
modes.  Of these, the last 4 modes have relatively low statistics in
planned $\ee$ collider experiments.  However, the HL-LHC is expected
to make high-precision measurements of the ratios of the
$\gamma\gamma$, $Z\gamma$, $\mu^+\mu^-$, and $ZZ^*$ branching ratios.
These ratios are especially favorable to measure in the hadronic
environment, since the four final states are measured in the dominant
central-rapidity production channel $h\to gg$, and the systematic
error from the production cross section can be arranged to cancel to a
great extent.   Combining this information, we have 6 high-precision
measurements from $\ee$ and 3 from HL-LHC.   The 20-variable system is
now closed.  It can be checked that there are no unexpected flat
directions in the determination of the SMEFT parameters from these
inputs.

Measurements in addition to these overdetermine the fit, and their
consistency provides useful cross-checks.   In particular, the
measurement of the $\sigma\times BR$ for the final state $ZZ^*$
provides
useful additional information,   This addition is especially important 
for the high-luminosity circular $\ee$ colliders.
On the other hand, the beam polarization asymmetry of the $\ee\to Zh$
cross section turns out to be exceptionally sensitive to the parameter
$c_{WW}$.  This gives an important input to the fit at linear
colliders with beam polarization~\cite{missed}. In practice, the two advantages
balance to a great extent, predicting similar performance for all four
of the currently proposed Higgs factories~\cite{compare}.  
Measurements of $\sigma\times BR$ values from the $W$-fusion reaction
$\ee\to \nu\bar \nu h$ can provide additional independent inputs, especially at
energies well above 250~GeV, that further constrain the SMEFT fit.

It is worth saying more about the role of the
decays $h\to WW^*$ and $h\to ZZ^*$ in this analysis.  Note that, at
the level of dimension-6 operators, there are no terms beyond those in
\leqn{Lsixone} and \leqn{Lsixtwo} that shift the $hWW$ and $hZZ$
couplings from their Standard Model values.   The SMEFT Lagragian
generates both of the tensor structures shown in \leqn{LhZZ} for the
$hZZ$ coupling, and two similar terms for the $hWW$ coupling.  Both
coefficients are allowed to take different values for $W$ and $Z$, but
the differences between these values are constrained by $SU(2)\times
U(1)$ symmetry.  In particular,
\beq
\eta_W =  - \half c_H  \ , \qquad   \eta_Z = -\half c_H - c_T  \ ,
\eeq{etarelation}
so the difference between the coefficients of the first tensor
structure is constrained by the precision electroweak constraints on
the $T$ parameter.  Further,
\beq
   \zeta_W =  (8c_{WW}) \ , \qquad  \zeta_Z = c_w^2 (8c_{WW}) + 2
   s_w^2 (8c_{WB}) + {s_w^4\over c_w^2} (8c_{BB}) \ ,
   \eeq{zetarelation}
 so these parameters cannot be very different if precision electroweak,
$\ee\to WW$, and $h\to \gamma\gamma$  measurements constrain the sizes
of   $c_{WB}$ and $c_{BB}$.    This is the reason that a
high-precision measurement of $BR(h\to ZZ^*)$ is not essential for the
success of the SMEFT fit.

\section{Prospects for precision Higgs boson measurements}
\label{sec:ILC}

Now that I have explained the mechanics of the SMEFT fit to the
projected results from $\ee$ Higgs factories, I would like to present
the sensitivities predicted for the measurement of Higgs boson
couplings.   Here I will present the results for ILC presented in
\cite{ILCprec}.
Similar results
are expected for any of the four Higgs factory proposals currently
under discussion~\cite{compare}.  

The proposed ILC program has two stages, the first at 250~GeV, the
second at 500~GeV in the center of mass.  In principle, a third stage
at 1~TeV is also possible with the same technology, either with a
longer tunnel or with high-gradient accelerating cavities that might
be available in the future~\cite{ILC}.

%%%%%%%%%%%%%%%%%%%%%%%%%%%%%%%%%%%%%%%%%%%%5
\begin{table}[h!]
\begin{center}
\begin{tabular}{l|cc|cc|cc}
      &  \multicolumn{2}{c}{ ILC250 }     &  
\multicolumn{2}{c}{ ILC500}  &
               \multicolumn{2}{c}{ ILC1000 }\\ 
coupling & full & no BSM & full & no BSM & full & no BSM \\ \hline 
$hZZ$            &             0.48 & 0.38 &            0.35 &  0.20    &
                 0.34  &  0.16   \\ 
$hWW$            &          0.48 & 0.38  &            0.35  &  0.20      &                                                
                              0.34 &  0.16 \\ 
 $hbb$            &     0.99  & 0.80  & 0.58 &0.43   &  0.47 & 0.31 \\ 
$h\tau\tau$    &          1.1 & 0.95 &   0.75 & 0.64 & 0.63 & 0.52 \\ 
$hgg$                  &  1.6 &  1.6    &   0.96  & 0.92    
 & 0.67 &  0.59 \\ 
$hcc$         &   1.8  &  1.8&  1.2   &   1.1 &    0.79 &   0.72  \\ 
$h\gamma\gamma$ &  1.1  & 1.1 &   1.0  &  0.97 & 0.94  & 0.89 \\ 
$h\gamma Z$     &  8.9 &  8.9 &   6.5 &  6.5 &  6.4  & 6.4 \\
$h\mu\mu$ &  4.0  & 4.0 &  3.8  &  3.8 &     3.4  &  3.4     \\ 
$htt$  &   ---     &   --- &   6.3 & 6.3  &      1.6 & 1.6    \\ 
$hhh$  &  ---    &  ---& 27 &  27 &    10  & 10  \\ \hline 
$\Gamma_{tot}$ & 2.3 & 1.3 & 1.6 & 0.70 &  1.4  &  0.50\\  
$\Gamma_{inv}$ &   0.36 & ---   & 0.32 & --- &  0.32  & --- \\  \hline
\end{tabular}
\end{center}
\caption{Projected uncertainties in the Higgs
  boson couplings for the ILC250, ILC500, and ILC1000, with
  precision LHC input~\cite{ILCprec}.  All values are  relative
  errors, in percent (\%).   The columns labelled ``full'' refer to
  the 
  22-parameter fit including the possibility of invisible and exotic
  Higgs boson decays.   The columns labelled ``no BSM'' refer to the
  20-parameter fit including only decays modes present in the SM. }
\label{tab:ILCresults}
\end{table}
%%%%%%%%%%%%%%%%%%%%%%%%%%%%%%%%%%%%%%%%%%%%%%%%%%%%

The analysis put forward by the ILC group includes one more possible
type of deviation from the Standard Model.  There may 
exist new particles with masses much lighter than $m_h$, perhaps
associated with a dark matter sector.   These can lead to invisible or
partially invisible decays of the Higgs boson~\cite{exotic}.   These
decays are actually observable using the reaction $\ee\to Zh$.  For example,
an invisible decay of the Higgs boson is indicated by an event with a
$Z$ boson at 110~GeV in the lab and nothing else.   In the ILC
analysis, 2 extra parameters are included, one representing the
branching ratio for fully invisible decays and and one representing
the branching ratio for other exotic decays that do not fit into any
preassigned category.  Since the invisible decay rate of the Higgs
boson is measured or bounded, these two parameters can be added to the
SMEFT fit without affecting its closure.   It is assumed that the loop
effects of the light particles do not affect precision electroweak
observables.  This is typically true for models of light dark matter
particles.

\begin{figure}
\begin{center}
\includegraphics[width=0.90\hsize]{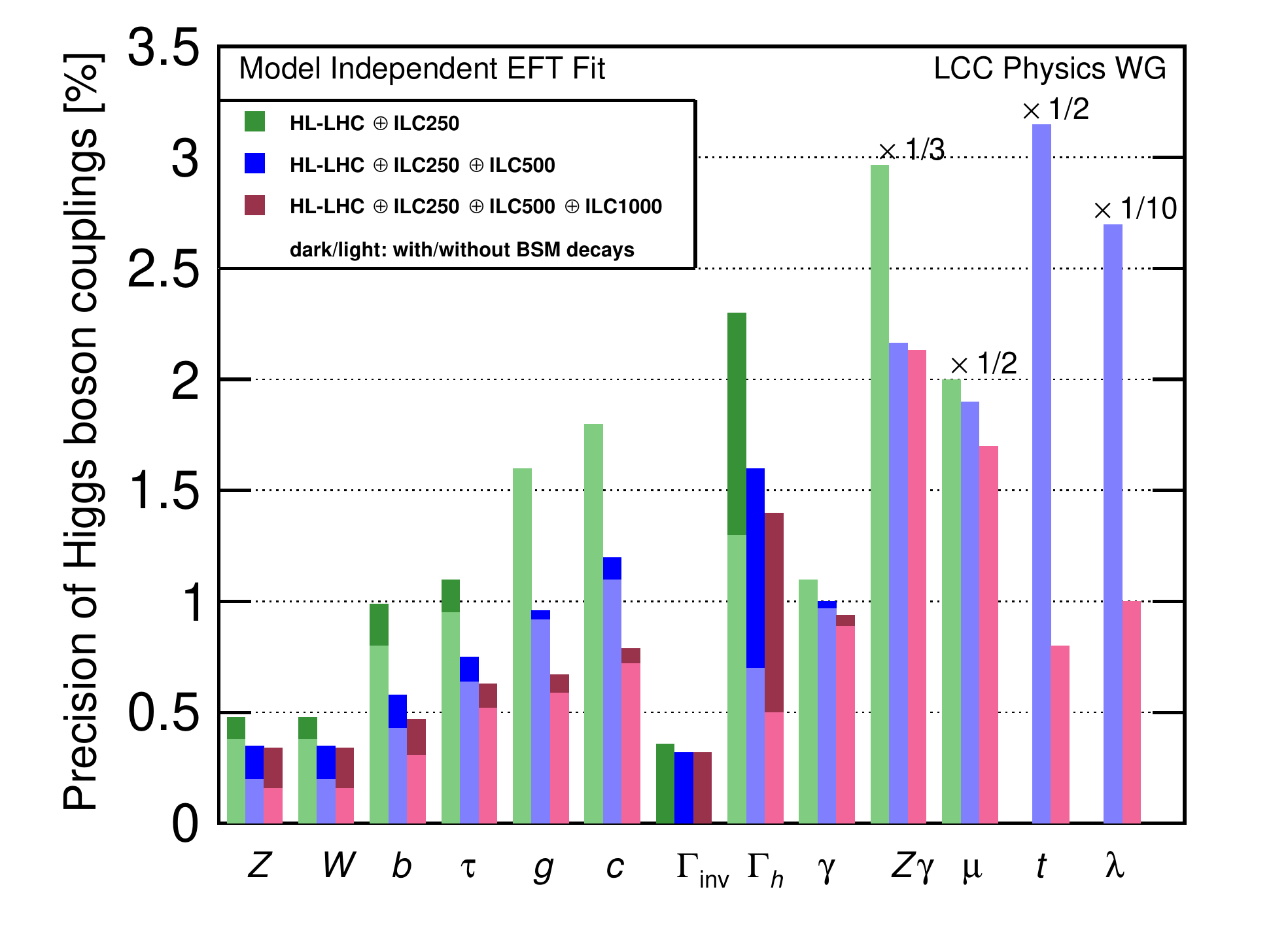}
\end{center}
\caption{Expected precision of the determination of Higgs boson
  coupling constants at the ILC, after its 250~GeV, 500~GeV, and
  1000~GeV stages~\cite{ILCprec}.   Input from specific measurements
  at  the HL-LHC is included, as described in the text.  Precisions
  are given in $\%$, but for the last four couplings, the estimates
  are rescaled by the factors shown in the figure.  The full column
  heights show the estimates  for an analysis that allows exotic Higgs boson
  decays.  When it is assumed that there are no exotic decays, the
  estimates improve to the heights shown by the light-colored bands. }
\label{fig:Higgscouplings}
\end{figure}

The results of this 22-parameter fit are shown in
Table~\ref{tab:ILCresults}, taken from \cite{ILCprec}.   Already at
the 250~GeV stage of the ILC, the
SMEFT fit with the expected experimental precision gives
uncertainties  less than 1\% for the important Higgs boson couplings
to $W$, $Z$, and $b$.
 As data is added at higher energies, using the
measurements of the  independent
$\ee\to \nu\bar\nu h$ reaction. the uncertainties on the $c$, $\tau$,
$g$, and (with the help of HL-LHC data) $\gamma$ couplings also reach
the 1\% level of accuracy.    Running at 1~TeV would further improve
these determinations, and also would bring the uncertainties on the
$t$ coupling and the Higgs self-coupling to 1.6\% and 10\%,
respectively.   Completing this program would give us experimental
determinations of the full suite of Higgs boson couplings, at a level
at which the possible effects of new physics would be expected 
to appear with high significance.   The improvement in the Higgs
coupling determinations expected from this program is shown
graphically in Fig.~\ref{fig:Higgscouplings}. 

%%%%%%%%%%%%%%%%%%%%%%%%%%%%%%%%%%%%%%%%%%%%%%%%%%%%%%%%%%%%%%%%%%%%%%%%%
%%%%%%%%%%%%%%%%%%%%%%%%%%%%%%%%%%%%%%%%%%%%%%%%%%%%%%%%%%%%%%%%%%%%%%%%%%%

\section{Conclusions}

Though the Standard Model is very successful in explaining current
experimental data, we should not claim that it is the final theory
of the fundamental interactions.  It is easy to call out phenomena in
the universe that the Standard Model does not account for. In this
article, I have emphasized
the mysteries surrounding the most important conceptual feature of 
the  Standard Model physics, the
spontaneous breaking of its gauge symmetry, which the model can
parametrize but which it is incapable of 
explaining.  To provide that explanation,
there must be new interactions lying undiscovered at higher energies.
The discovery of these new interactions will be as consequential as
the discovery of the Standard Model itself.

Among the methods of searching for new interactions, I have emphasized
here the precision study of electroweak and Higgs interactions.  A
useful tool for understanding the implications of our current level of
precision and the significance of future improvements is the Standard
Model Effective Field Theory.  In this article, I have reviewed the
application of SMEFT to a variety of experimental measurements,
including the future program of precision measurements of the
couplings of the Higgs boson.

I have argued that it is within our current technical capabilities to
measure the couplings of the Higgs boson with a precision of 1\% or
below.  This  is not only a matter
of completing the verification of the Standard Model.  If
we can reach this high level of precision in the study of the Higgs
boson, we will have the possibility of observing with high confidence
the characteristic effects of new interactions that could explain the
origin of  Higgs electroweak symmetry breaking.  This study
could well be the one that breaks through to the next level of
fundamental physics, the level that answers the questions that seem
intractable today.

We should not miss this opportunity to move to the next deeper level of
our understanding of fundamental physics.

\Acknowledgements

I am grateful to  Bryan Lynn, Harsh Mathur, Glenn Starkman, Kellen
McGee,
and their team at
Case Western Reserve for inviting me to this symposium, and for making
it so memorable. I thank Tim Barklow, Christophe Grojean, Sunghoon
Jung, Jenny List, and Junping Tian for leading me into the world of
SMEFT
Higgs analysis.  I am grateful to many
other colleagues for enhancing my understanding of this formalism.
This work was supported by
the
U.S. Department 
of Energy under contract DE--AC02--76SF00515.


\begin{thebibliography}{99}

%%
%%  bibliographic items can be constructed using the LaTeX format in SPIRES:
%%    see    http://www.slac.stanford.edu/spires/hep/latex.html
%%  SPIRES will also supply the CITATION line information; please include it.
%%

\bibitem{Brivio}
   I.~Brivio and M.~Trott,
  %``The Standard Model as an Effective Field Theory,''
  Phys.\ Rept.\  {\bf 793}, 1 (2019)
%  doi:10.1016/j.physrep.2018.11.002
  [arXiv:1706.08945 [hep-ph]].


\bibitem{LG}
V. L. Ginzburg and L. D. Landau, Zh. Eksp. Theor. Fiz {\bf 20}, 1064 (1950).

\bibitem{BCS}
 J. Bardeen, L. N. Cooper, and J. R. Schrieffer, Phys. Rev. {\bf 106},
 162, {\bf 108}, 1175 (1957). 


\bibitem{Wilson}
K.~G.~Wilson,
%``Nonlagrangian models of current algebra,''
Phys.\ Rev.\  \textbf{179}, 1499 (1969).
%doi:10.1103/PhysRev.179.1499
%



\bibitem{Weinberg}
S.~Weinberg,
%``Phenomenological Lagrangians,''
Physica A \textbf{96}, 327 (1979).
%doi:10.1016/0378-4371(79)90223-1



\bibitem{GasserL}
  J.~Gasser and H.~Leutwyler,
%``Chiral Perturbation Theory to One Loop,''
Annals Phys.\  \textbf{158}, 142 (1984),
% doi:10.1016/0003-4916(84)90242-2
%``Chiral Perturbation Theory: Expansions in the Mass of the Strange Quark,''
Nucl.\ Phys.\ B \textbf{250}, 465 (1985).
%doi:10.1016/0550-3213(85)90492-4

\bibitem{WeinbergSM}
S.~Weinberg,
%``Nonabelian Gauge Theories of the Strong Interactions,''
Phys.\ Rev.\ Lett.\  \textbf{31}, 494 (1973).
%doi:10.1103/PhysRevLett.31.494



  \bibitem{Nanopoulos}
D.~V.~Nanopoulos,
%``Towards a renormalizable unified gauge theory of strong, electromagnetic and weak interactions,''
Lett.\ Nuovo Cim.\  \textbf{8}, 873 (1973).
% doi:10.1007/BF02727401


\bibitem{Restimate}
  There is an extensive literature on the estimation of SMEFT
    coefficients.  See, for example, \cite{Rattazzi1,Rattazzi2,Rattazzi3}.

\bibitem{Rattazzi1}
G.~Giudice, C.~Grojean, A.~Pomarol and R.~Rattazzi,
%``The Strongly-Interacting Light Higgs,''
JHEP \textbf{06}, 045 (2007)
%doi:10.1088/1126-6708/2007/06/045
[arXiv:hep-ph/0703164 [hep-ph]].

\bibitem{Rattazzi2}
I.~Low, R.~Rattazzi and A.~Vichi,
%``Theoretical Constraints on the Higgs Effective Couplings,''
JHEP \textbf{04}, 126 (2010)
%doi:10.1007/JHEP04(2010)126
[arXiv:0907.5413 [hep-ph]].



\bibitem{Rattazzi3}
D.~Liu, A.~Pomarol, R.~Rattazzi and F.~Riva,
%``Patterns of Strong Coupling for LHC Searches,''
JHEP \textbf{11}, 141 (2016)
%doi:10.1007/JHEP11(2016)141
[arXiv:1603.03064 [hep-ph]].

\bibitem{Warsaw} 
  B.~Grzadkowski, M.~Iskrzynski, M.~Misiak and J.~Rosiek,
  %``Dimension-Six Terms in the Standard Model Lagrangian,''
  JHEP {\bf 1010}, 085 (2010)
  %doi:10.1007/JHEP10(2010)085
  [arXiv:1008.4884 [hep-ph]].

  \bibitem{Alonso} 
  R.~Alonso, E.~E.~Jenkins, A.~V.~Manohar and M.~Trott,
  %``Renormalization Group Evolution of the Standard Model Dimension Six Operators III: Gauge Coupling Dependence and Phenomenology,''
  JHEP {\bf 1404}, 159 (2014)
 % doi:10.1007/JHEP04(2014)159
  [arXiv:1312.2014 [hep-ph]].
  %%CITATION = doi:10.1007/JHEP04(2014)159;%%

  \bibitem{Abolins}
M.~Abolins, \etal,
% ``Testing the Compositeness of Quarks and Leptons,''
in {\it Proceedings of the 
1982 DPF Summer Study on Elementary Particle Physics and Future
Facilities (Snowmass 82)},  R. Donaldson, R. Gustafson, and F. Paige, eds.
eConf \textbf{C8206282}, 274 (1982).

\bibitem{ELP}
E.~Eichten, K.~D.~Lane and M.~E.~Peskin,
%``New Tests for Quark and Lepton Substructure,''
Phys.\ Rev.\ Lett.\  \textbf{50}, 811 (1983).
%doi:10.1103/PhysRevLett.50.811

\bibitem{Bourilkov}
  D.~Bourilkov,
%``Search for TeV strings and new phenomena in Bhabha scattering at LEP-2,''
Phys.\ Rev.\ D \textbf{62}, 076005 (2000)
%doi:10.1103/PhysRevD.62.076005
[arXiv:hep-ph/0002172 [hep-ph]].

\bibitem{ATLAScontact}
  M.~Aaboud {\it et al.} [ATLAS Collaboration],
  %``Search for new phenomena in dijet events using 37 fb$^{-1}$ of $pp$ collision data collected at $\sqrt{s}=$13 TeV with the ATLAS detector,''
  Phys.\ Rev.\ D {\bf 96}, 052004 (2017)
 % doi:10.1103/PhysRevD.96.052004
  [arXiv:1703.09127 [hep-ex]].
  %%CITATION = doi:10.1103/PhysRevD.96.052004;%%



\bibitem{CMScontact}
 A.~M.~Sirunyan {\it et al.} [CMS Collaboration],
  %``Search for new physics in dijet angular distributions using proton–proton collisions at $\sqrt{s}=$ 13 TeV and constraints on dark matter and other models,''
  Eur.\ Phys.\ J.\ C {\bf 78}, 789 (2018)
  %doi:10.1140/epjc/s10052-018-6242-x
  [arXiv:1803.08030 [hep-ex]].

\bibitem{LPS}
   B.~W.~Lynn, M.~E.~Peskin and R.~G.~Stuart,
   % ``RADIATIVE CORRECTIONS IN SU(2) x U(1): LEP / SLC,''
   in {\it Physics at LEP}, J. Ellis and R. Peccei, eds.  CERN Yellow
   Report 86-02 (1986).

 \bibitem{PT}
 M.~E.~Peskin and T.~Takeuchi,
%``A New constraint on a strongly interacting Higgs sector,''
Phys.\ Rev.\ Lett.\  \textbf{65}, 964 (1990), 
%doi:10.1103/PhysRevLett.65.964
%``Estimation of oblique electroweak corrections,''
Phys.\ Rev.\ D \textbf{46}, 381 (1992).
%doi:10.1103/PhysRevD.46.381

\bibitem{SSVZ}
P.~Sikivie, L.~Susskind, M.~B.~Voloshin and V.~I.~Zakharov,
%``Isospin Breaking in Technicolor Models,''
Nucl.\ Phys.\ B \textbf{173}, 189 (1980).
%doi:10.1016/0550-3213(80)90214-X


  

\bibitem{LEPEWZ}
 S.~Schael {\it et al.} [ALEPH and DELPHI and L3 and OPAL and SLD Collaborations and LEP Electroweak Working Group and SLD Electroweak Group and SLD Heavy Flavour Group],
  %``Precision electroweak measurements on the $Z$ resonance,''
  Phys.\ Rept.\  {\bf 427}, 257 (2006)
 % doi:10.1016/j.physrep.2005.12.006
  [hep-ex/0509008].

\bibitem{Gfitter}
M.~Baak \textit{et al.} [Gfitter Group],
%``The global electroweak fit at NNLO and prospects for the LHC and ILC,''
Eur.\ Phys.\ J.\ C \textbf{74}, 3046 (2014)
%doi:10.1140/epjc/s10052-014-3046-5
[arXiv:1407.3792 [hep-ph]].

\bibitem{ATLASh}
G.~Aad \textit{et al.} [ATLAS Collaboration],
%``Observation of a new particle in the search for the Standard Model Higgs boson with the ATLAS detector at the LHC,''
Phys.\ Lett.\ B \textbf{716}, 1 (2012)
%doi:10.1016/j.physletb.2012.08.020
[arXiv:1207.7214 [hep-ex]].



  \bibitem{CMSh}
S.~Chatrchyan \textit{et al.} [CMS Collaboration],
%``Observation of a New Boson at a Mass of 125 GeV with the CMS Experiment at the LHC,''
Phys.\ Lett.\ B \textbf{716}, 30 (2012)
%doi:10.1016/j.physletb.2012.08.021
[arXiv:1207.7235 [hep-ex]].
%

\bibitem{ATLAShm}
  M.~Aaboud {\it et al.} [ATLAS Collaboration],
  % ``Measurement of the Higgs boson mass in the $H\rightarrow ZZ^* \rightarrow 4\ell$ and $H \rightarrow\gamma\gamma$ channels with $\sqrt{s}=13$ TeV $pp$ collisions using the ATLAS detector,''
  Phys.\ Lett.\ B {\bf 784}, 345 (2018)
%  doi:10.1016/j.physletb.2018.07.050
  [arXiv:1806.00242 [hep-ex]].


\bibitem{CMShm}
 A.~M.~Sirunyan {\it et al.} [CMS Collaboration],
 % ``Measurements of properties of the Higgs boson decaying into the
 % four-lepton final state in pp collisions at $ \sqrt{s}=13 $ TeV,''
  JHEP {\bf 1711}, 047 (2017)
 % doi:10.1007/JHEP11(2017)047
  [arXiv:1706.09936 [hep-ex]].
  %%CITATION = doi:10.1007/JHEP11(2017)047;%%

\bibitem{PDGHiggs}
M. Tanabashi \etal\  (Particle Data Group), Phys. Rev. {\bf D 98},
  030001 (2018),
  and 2019 update.

\bibitem{CMScomb}
 A.~M.~Sirunyan {\it et al.} [CMS Collaboration],
  %``Combined measurements of Higgs boson couplings in proton–proton collisions at $\sqrt{s}=13\,\text {Te}\text {V} $,''
  Eur.\ Phys.\ J.\ C {\bf 79}, 421 (2019)
%  doi:10.1140/epjc/s10052-019-6909-y
  [arXiv:1809.10733 [hep-ex]].
  %%CITATION = doi:10.1140/epjc/s10052-019-6909-y;%%


  \bibitem{ATLAScomb}
 G.~Aad {\it et al.} [ATLAS Collaboration],
  %``Combined measurements of Higgs boson production and decay using up to $80$ fb$^{-1}$ of proton-proton collision data at $\sqrt{s}=$ 13 TeV collected with the ATLAS experiment,''
  Phys.\ Rev.\ D {\bf 101}, 012002 (2020)
%  doi:10.1103/PhysRevD.101.012002
  [arXiv:1909.02845 [hep-ex]].
  %% CITATION = doi:10.1103/PhysRevD.101.012002;%%


\bibitem{WellsZhang}
J.~D.~Wells and Z.~Zhang,
%``Effective field theory approach to trans-TeV supersymmetry: covariant matching, Yukawa unification and Higgs couplings,''
JHEP \textbf{05}, 182 (2018)
%doi:10.1007/JHEP05(2018)182
[arXiv:1711.04774 [hep-ph]].


\bibitem{myCERNSS}
M.~E.~Peskin, in {\it Proceedings of the 2016 European School of
  High-Energy
  Physics}, M. Mulders and G. Zanderighi, eds.  CERN Yellow Report
CERN-2017-009-SP (2017)
%``Lectures on the Theory of the Weak Interaction,''
%doi:10.23730/CYRSP-2017-005.1
[arXiv:1708.09043 [hep-ph]].



\bibitem{Barklow1}
T.~Barklow, K.~Fujii, S.~Jung, R.~Karl, J.~List, T.~Ogawa, M.~E.~Peskin and J.~Tian,
%``Improved Formalism for Precision Higgs Coupling Fits,''
Phys.\ Rev.\ D \textbf{97}, 053003 (2018)
%doi:10.1103/PhysRevD.97.053003
[arXiv:1708.08912 [hep-ph]].


\bibitem{Barklow2}
T.~Barklow, K.~Fujii, S.~Jung, M.~E.~Peskin and J.~Tian,
%``Model-Independent Determination of the Triple Higgs Coupling at e+e- Colliders,''
Phys.\ Rev.\ D \textbf{97}, 053004 (2018)
%doi:10.1103/PhysRevD.97.053004
[arXiv:1708.09079 [hep-ph]].

\bibitem{HLLHCproj}
 M.~Cepeda, \etal, in {\it Physics of the HL-LHC, and perspectives of
  the HE-LHC}, A.~Dainese, M.~Mangano, A.~B.~Meyer, A.~Nisati,
G.~Salam and M.~Vesterinen, eds. 
%``Report from Working Group 2,''
CERN Yellow Report   CERN-2019-007  (2019)
%doi:10.23731/CYRM-2019-007.221
[arXiv:1902.00134 [hep-ph]].



\bibitem{CEPC1}
CEPC Study Group,  IHEP Report IHEP-CEPC-DR-2018-01 (2018) 
%``CEPC Conceptual Design Report: Volume 1 - Accelerator,''
[arXiv:1809.00285 [physics.acc-ph]].


\bibitem{CEPC2}
  J.~B.~Guimarães da Costa \textit{et al.} [CEPC Study Group],
  IHEP Report IHEP-CEPC-DR-2018-02 (2018) 
%``CEPC Conceptual Design Report: Volume 2 - Physics & Detector,''
[arXiv:1811.10545 [hep-ex]].


\bibitem{FCCee}
A.~Abada \textit{et al.} [FCC Collaboration],
%``FCC-ee: The Lepton Collider,''
Eur.\ Phys.\ J.\ ST \textbf{228}, 261  (2019),
	CERN-ACC-2018-0057.
%doi:10.1140/epjst/e2019-900045-4


\bibitem{ILC}
P.~Bambade, \etal,
%``The International Linear Collider: A Global Project,''
arXiv:1903.01629 [hep-ex].

  

\bibitem{ILCprec}
  K.~Fujii \textit{et al.} [LCC Physics Working Group],
%``Tests of the Standard Model at the International Linear Collider,''
  arXiv:1908.11299 [hep-ex].


\bibitem{CLIC}
P.~Burrows \textit{et al.} [CLICdp and CLIC],
%``The Compact Linear Collider (CLIC) - 2018 Summary Report,''
CERN Yellow Report  CERN-2018-005  (2018)
%doi:10.23731/CYRM-2018-002
[arXiv:1812.06018 [physics.acc-ph]].

\bibitem{missed}
  This point, made quite explicitly in \cite{Barklow1},
  seems to have been missed in the comparisons of
    linear and circular colliders given in \cite{FCCee} and other
    reports from the FCC-ee group.

  \bibitem{compare}
    These comparisons are shown  in detail in the tables of \cite{ILC}, Section~11 ,
    and in those from the SMEFT analysis presented in \cite{deBlas}.

 \bibitem{deBlas}
J.~de Blas, \etal\  [Higgs @ Future Colliders Working Group],
%``Higgs Boson Studies at Future Particle Colliders,''
arXiv:1905.03764 [hep-ph].
 
\bibitem{exotic}
D.~Curtin, \etal, 
%``Exotic decays of the 125 GeV Higgs boson,''
Phys.\ Rev.\ D \textbf{90},  075004 (2014)
%doi:10.1103/PhysRevD.90.075004
[arXiv:1312.4992 [hep-ph]].

\end{thebibliography}
\end{document}